\documentclass[aps,prb,longbibliography,a4paper,twocolumn,superscriptaddress,citeautoscript]{revtex4-1}
\usepackage[normalem]{ulem}
\usepackage{hyperref}
\usepackage{color}
\usepackage[english]{babel}
\usepackage{lmodern}
\usepackage[T1]{fontenc}
\usepackage{amsmath}
\usepackage{units}
\usepackage[pdftex]{graphicx}
\usepackage{grffile}
\usepackage{color}
\usepackage[bottom]{footmisc}
\usepackage{graphicx}
\usepackage{dcolumn}
\usepackage{bm}
\usepackage{soul}
\begin{document}
\title{Magnetic field tuning of the valley population in the Weyl phase of Nd$_2$Ir$_2$O$_7$}
\author{Itzik Kapon}
\email{itzhak.kapon@unige.ch}
\affiliation{Department of Quantum Matter Physics, University of Geneva, 24  Quai Ernest-Ansermet, 1211 Geneva, Switzerland}
\author{Carl Willem Rischau}
\affiliation{Department of Quantum Matter Physics, University of Geneva, 24  Quai Ernest-Ansermet, 1211 Geneva, Switzerland}
\author{Bastien Michon}
\affiliation{Department of Quantum Matter Physics, University of Geneva, 24  Quai Ernest-Ansermet, 1211 Geneva, Switzerland}
\author{Kai Wang}
\affiliation{Department of Quantum Matter Physics, University of Geneva, 24  Quai Ernest-Ansermet, 1211 Geneva, Switzerland}
\author{Bing Xu}
\affiliation{Department of Physics and Fribourg Center for Nanomaterials, University of Fribourg, Fribourg, Switzerland}
\author{Qiu Yang}
\affiliation{Department of Physics, University of Tokyo, Hongo, Bunkyo-ku, Tokyo 113-0033, Japan}
\author{Satoru Nakatsuji}
\affiliation{Department of Physics, University of Tokyo, Hongo, Bunkyo-ku, Tokyo 113-0033, Japan}
\affiliation{Institute for Solid State Physics, University of Tokyo, Kashiwa, Chiba 277-8581, Japan}
\affiliation{Institute for Quantum Matter and Department of Physics and Astronomy, Johns Hopkins University, Baltimore, Maryland 21218, USA} 
\affiliation{Trans-scale Quantum Science Institute, University of Tokyo, Hongo, Bunkyo-ku, Tokyo 113-8654, Japan}
\author{Dirk van der Marel}
\affiliation{Department of Quantum Matter Physics, University of Geneva, 24  Quai Ernest-Ansermet, 1211 Geneva, Switzerland}
\date{\today}
\begin{abstract}
The frustrated magnet Nd$_2$Ir$_2$O$_7$, where strong correlations together with spin-orbit coupling play a crucial role, is predicted to be a Weyl semimetal and to host topological pairs of bulk Dirac-like valleys.
Here we use an external magnetic field to manipulate the localized rare earth 4f moments coupled to the 5d electronic bands.
Low energy optical spectroscopy reveals that a field of only a few teslas suffices to create charge compensating pockets of holes and electrons in different regions of momentum space, thus introducing a valley population shift that can be tuned with the field.

\end{abstract}
\maketitle
\section{Introduction}
Dirac materials are understood today in terms of common topological properties, although they encompass vast range of compounds with different underlying physical mechanisms which bring upon the low energy Dirac dispersion~\cite{vafek2014dirac,armitage2018weyl}.
These materials, be that two (2D) or three (3D) dimensional, host Dirac cones in their electronic band structure, in the bulk or on the surface, where the conduction and valence bands are degenerate at a discrete set of points in the Brillouin zone, and disperse linearly around these points.
Graphene is the most famous 2D one, however many more materials belong to this family: Weyl semimetals like TaAs~\cite{lv2015experimental}, Mn$_3$Sn~\cite{nakatsuji2015large} and CoSi~\cite{rao2019observation}, the topological insulator Bi$_2$Se$_3$, for the spin-density wave state of BaFe$_2$As$_2$~\cite{richard2010observation} and it has been predicted for the pseudogap phase of the cuprate high-$T_c$ superconductors~\cite{borne2010specific}, to only name a few.
Recently, it has been proposed to use these materials for a new type of electronics utilizing the Dirac valleys, namely, ``valleytronics"~\cite{vitale2018valleytronics}.
The main challenge towards achieving this is valley polarization - the ability to selectively control the different Dirac cones with external forces, like circularly polarized light or magnetic field.

Here we show that the pyrochlore iridate Nd$_2$Ir$_2$O$_7$ is a new candidate for future valleytronics technologies.
It has been theoretically predicted to be a Weyl semimetal~\cite{WanTurner2011}, thus hosting distinguishable Dirac nodes.
The sister compound Pr$_2$Ir$_2$O$_7$ was experimentally reported to be a correlated Weyl semimetal under external strain~\cite{li2021correlated,ohtsuki2019strain}.
In our compound, the localized 4f electronic magnetic moments on the Nd sites interact with the itinerant 5d electrons from the Ir.
The Ir electrons carry an effective pseudospin $j=1/2$ due to spin-orbit interaction and crystal field splitting.
They order antiferromagnetically (AFM) at $T_{N}=37 $~K in the non-collinear all-in-all-out (AIAO) state with the Ir moments pointing inside or outside the tetrahedra, and create an exchange field of $\Delta_{0}=6.5 $~K on the Nd sites.
The Nd moments become detectable at $T\approx13 $~K and are also arranged in the AFM AIAO configuration.
An open question concerns the influence of the Nd moments on the Ir ones.
Moreover, the effect of a magnetic field could be amplified at the Ir sites by the strong Nd moments and the f-d interaction~\cite{tian2016field}.

We follow this approach, and use magneto-optical spectroscopy and magneto-calorimetry combined with mean-field calculations to show that magnetic field of seven tesla creates charge compensated pockets at the different Weyl nodes in the Brillouin zone.
This is due to the f-d interaction, where the Nd magnetic moments are re-oriented with the field and serve as field boosters at the Ir sites, thus altering the band structure.
Therefore, it is possible to tune the valley populations with modest fields.
It has also been suggested lately~\cite{wang2020unconventional} that this material demonstrates unconventional free charge, based on the quadratic in temperature dependence of the Drude spectral weight without magnetic field.
We comment on this interpretation and explain this phenomenon in the framework of charge compensated Weyl points.

\begin{figure*}[tbph!]
\includegraphics[width=2\columnwidth]{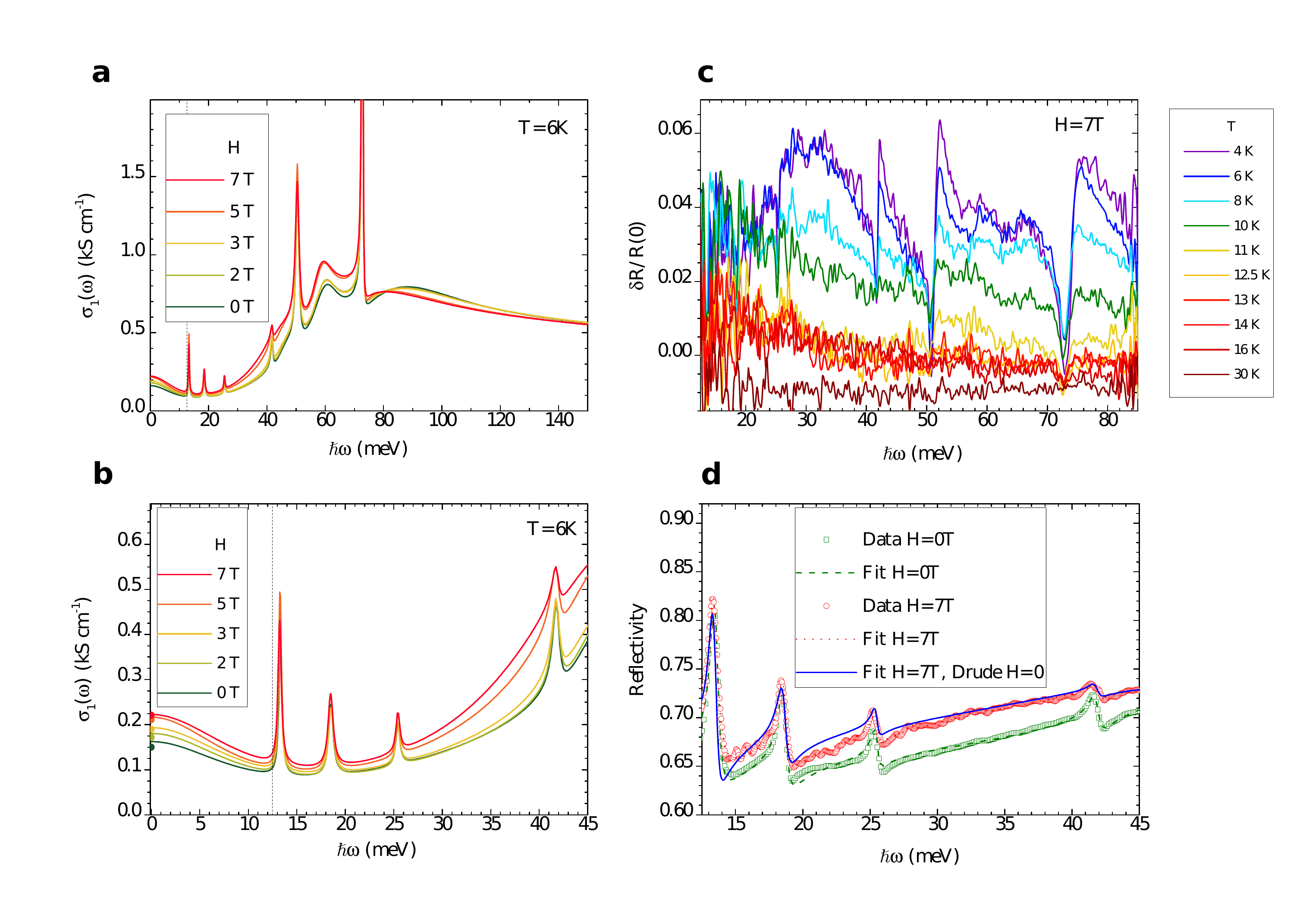}
\caption{Optical conductivity data. \textbf{a.} Real part of the optical conductivity $\sigma_1$ as function of energy for different magnetic fields at $T=6 $~K.
The main field effects are red shift of the spectra and the increase of the Drude component.
\textbf{b.} Same as (a) but for low energy.
DC conductivity data are presented as solid circles.
Vertical dashed lines (\textbf{a} and \textbf{b}) mark the energy below which the fits are extrapolated outside our measured data range.
\textbf{c.} Temperature evolution of the relative reflectivity at H=7T.
$R_0$ is the zero-field reflectivity and $\delta R = R(H)-R_0$.
Above $T\approx 13$~K we are not able to detect any change with field.
\textbf{d.} Fits to the reflectivity data at 0T (black) and 7T (red) with Drude part taken as constant with field (blue).}
\label{Sigma1Data}
\end{figure*}

\begin{figure}[tbph]
\includegraphics[width=\columnwidth]{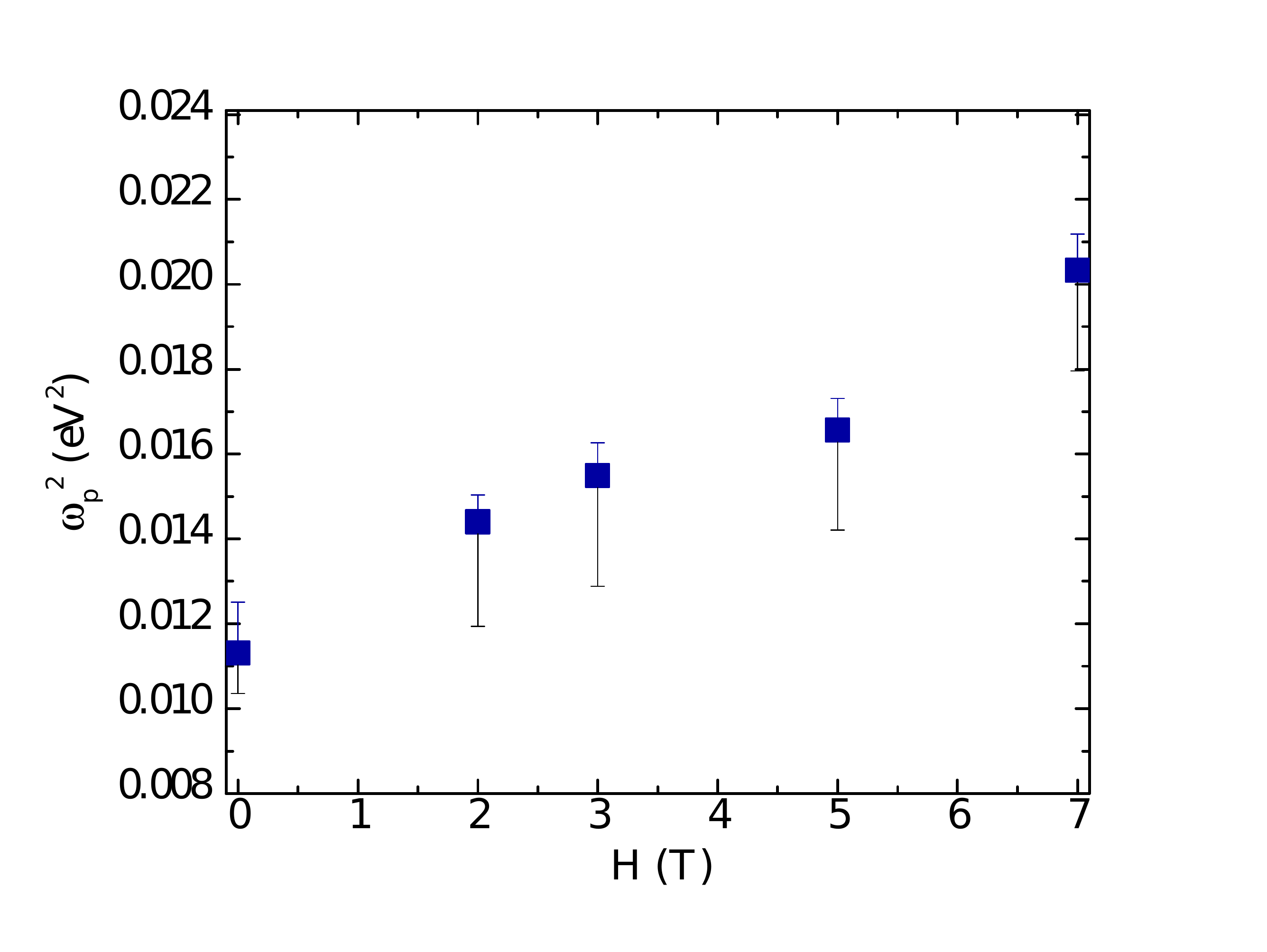}
\caption{Drude spectral Weight at $T=6 $~K. The free charge carrier density doubles from 0 to 7 tesla, supporting the creation of charge compensated pockets.
Antiferromagnetic domain walls exist until 2-3 tesla, as known from ~\cite{ma2015mobile}.
Error bars correspond to $1\%$ error of the reflectivity.}
\label{spectralweight}
\end{figure}

\begin{figure*}[tbph]
\includegraphics[width=2\columnwidth]{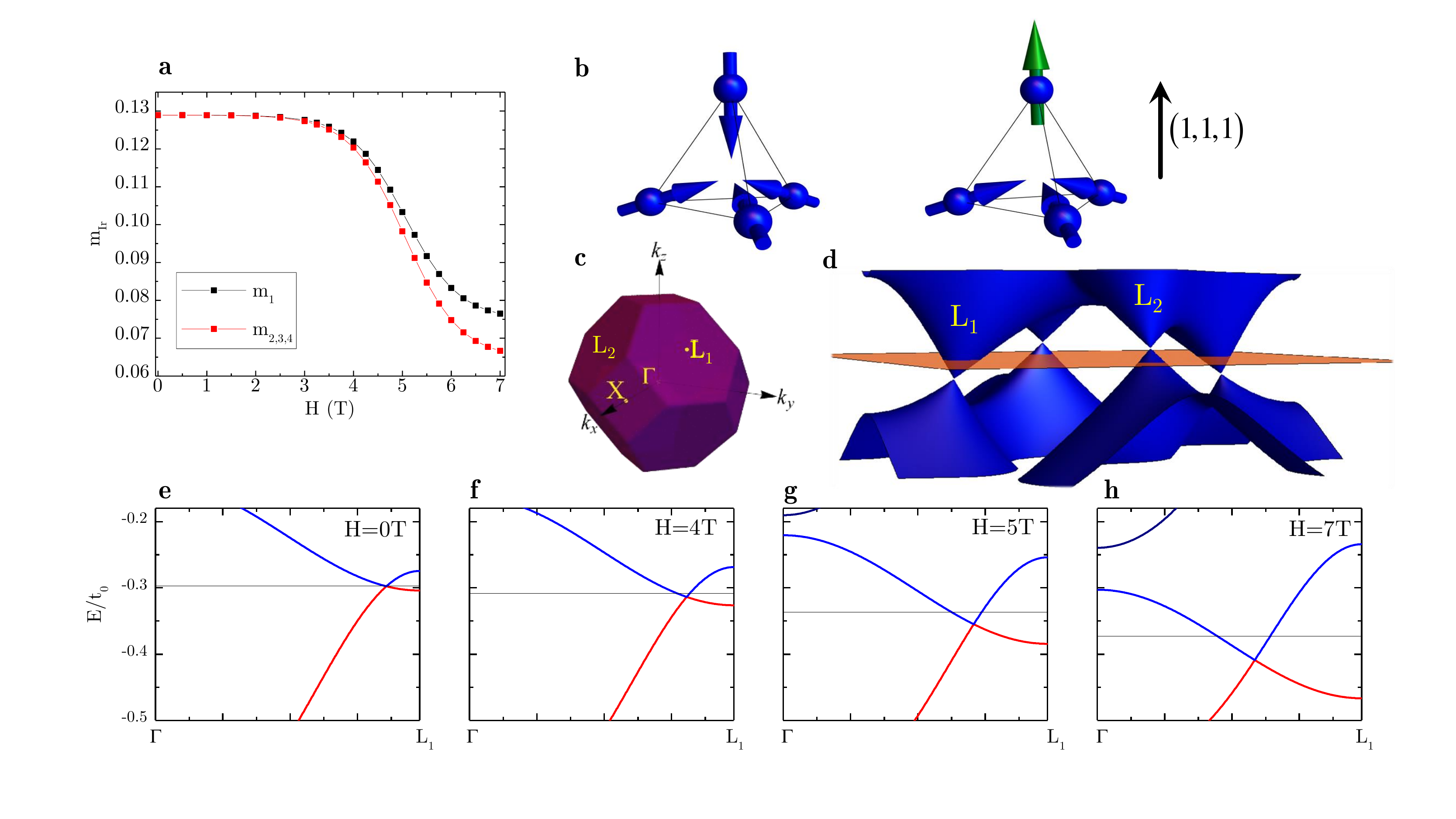}
\caption{Simulation data. \textbf{a)} Ir average magnetic moments as function of field.
Black squares  represent Ir pseudospin along the $(1,1,1)$ field direction, whereas red circles represent the other three Ir's.
 \textbf{b)} Illustration of the 4-in-0-out and 3-in-1-out configurations of the Nd moments.
The Nd atoms reside on the tetrahedron's corners.
\textbf{c)} The first Brillouin zone with high symmetry points marked.
\textbf{d)} Illustration of two pairs of Weyl points for $H=7 $~T in the $\Gamma L_{1}L_{2}$ plane.
Orange plane represents the chemical potential.
Small holes and electrons pockets are seen in the two nonequivalent points $L_1$ and $L_2$.
\textbf{e-h)} Band structure along $\Gamma L_{1}$ line for various fields, where a transition from Weyl semimetal to Weyl metal is clearly observed.}
\label{BandStructure}
\end{figure*}

\section{Experimental results}
For the present experiments we used the single crystal of Ref.~\onlinecite{wang2020unconventional}.
We measured reflectivity at temperatures between 4 to 300~K at magnetic fields from 0 to 7~T along the $(1,1,1)$ direction.
Optical conductivity spectra $\sigma_1(\omega)$ obtained from the reflectivity data (see Appendix~\ref{appendix:magneto-reflectance}) are shown for $T=6$~K for different magnetic fields in Fig.~\ref{Sigma1Data} (a) and (b).
As the field is increased, the conductivity is rising and red-shifting throughout the low energy range from 0 to 100 ~meV.
Specifically, the Drude part is increasing.
The strong spectral weight increase of the inter band transitions from 15-45 meV (see Fig. ~\ref{Sigma1Data}{\bf b}) brings up the question whether the increase in the dc conductivity can partly be attributed to the tail of the enhanced interband optical conductivity. When decomposing using the Drude-Lorentz model, by construction there is no interband contribution at zero frequency because of the mathematical properties of the Lorentz component. That said, in a microscopic calculation it is possible to have a non-zero interband conductivity at zero frequency, provided that two or more bands are overlapping and these bands cross right at the Fermi surface. However, we will see below that such a scenario is not supported by the electronic structure calculations.
Some peaks corresponding to phonons also exhibit an asymmetric change with the field.
To follow the temperature evolution at constant field, Fig.~\ref{Sigma1Data}(c) presents the relative change in the measured reflectivity $\delta R \left(H\right)/R\left(0\right) = \left(R\left(H\right) - R \left(0\right)\right)/R \left(0\right) $ at $H=7 $~T for temperatures from 4 to 30 $~K$.
Field response is only observed below $T\approx 13 $~K.
This is expected, as the Ir 5d electrons carry a small magnetic moment of approximately $0.2 \mu_B$ relative to the Nd 4f electrons with $2.4 \mu_B$ as measured by neutron scattering~\cite{tomiyasu2012neutrons}, where $\mu_B$ is the Bohr magneton.
Nonetheless, one must bear in mind that optical conductivity is mostly sensitive to the itinerant electrons, i.e., the Ir electronic bands.
Thus, it must be deduced that the external magnetic field indirectly influences the 5d bands via the 4f moments, whose coupling to the former comprises the biggest effect on their electronic phase.
See the computation and discussion sections for more details.

Fig.~\ref{spectralweight} presents the evolution of the Drude spectral weight with field, $\omega_{p}^2\left(H\right)$.
Going from zero to seven tesla the spectral weight doubles.
Thus, the free charge carrier density increases with field, as also observed in transport~\cite{ueda2015magnetoresistance,nakayama2016arpes}.
DC conductivity of our sample is presented as solid circles in Fig.~\ref{Sigma1Data}(b).
To further evaluate the free charge response to the field, we compare in Fig.~\ref{Sigma1Data}(d) the fits to the reflectivity data at $T=6 $~K for $H=0 $~T (green dashed line) and $H=7 $~T (red dotted line).
For the latter, we also present the fit while constraining the Drude part to its value at 0T (blue solid line).
The relative change in reflectivity due to the Drude part comprises a major contribution in the low energy regime.
We will later argue that this originates from valley population shift.

Next we describe our results from specific heat measurements under magnetic field.
The data (see Appendix~\ref{appendix:cv}) confirm that the most probable scenario is that above 2-3 tesla there is only one AF domain in our crystal.
It consists of Nd moments pointing towards the center of the tetrahedron, with three of them pointing along the favorable field direction, i.e., having positive projection along $(1,1,1)$.
That configuration is named ``4-in-0-out" (4-0) and illustrated in Fig.~\ref{BandStructure}(b) on the left panel.
This result is supported by a previous study that followed the evolution of domain walls in Nd$_2$Ir$_2$O$_7$~\cite{tardif2015domains}.
We extract the exchange field at the Nd sites exerted by the Ir moments as a function of magnetic field $\Delta_0 (H)$.
This effective potential will be used in our model for the f-d interaction (see Appendix~\ref{appendix:bands}).

\section{Model calculations}
To interpret the observed behavior, we conducted a Hartree mean field type calculation, following the tight binding Hamiltonian from Ref.~\onlinecite{witczak2013effectiveHamiltonian}, to which we added an effective f-d interaction.
To this end, we follow Chen and Hermele~\cite{chen2012fd}, starting with a general term describing coupling between Ir 5d electrons and Ising like Nd moments pointing in their local $z$ axis:

\begin{equation}
H_{fd}=\sum_{\left\langle i,j \right\rangle}\sum_{j\in Nd} \sum_{i\in Ir} \hat{\tau }_{j}^{z} \left[ \left( d_{i\alpha }^{\dagger }\frac{\pmb{\sigma}_{\alpha \beta }}{2}{d}_{i\beta }  \right) \cdot \mathbf{v}_{ji}  \right],
\label{eq:Hfd}
\end{equation}
where $\hat{\tau }_{j}^{z}$ is a Pauli matrix representing the j-th Nd Ising spin, $d_{i\alpha }^{\dagger}$ is a creation operator of 5d electron on the i-th Ir site with pseudo-spin $\alpha$, $\pmb{\sigma}$ is a vector of Pauli matrices, and $\mathbf{v}_{ji}$ are vectors describing the symmetry allowed interaction, and contain two parameters $c_1$ and $c_2$.
Next, we substitute $\hat{\tau }_{j}^{z}$ with an average magnetic moment $M_{j}^{z}$, as described hereafter.

Using our specific heat data, we construct an expression for the effective potential acting on the 4f moments.
Taking into account the Zeeman term, it is given by $\Delta _{j}^{'}(T,H)=\Delta_{0}(H)+ \alpha g \mu_B H$, where $\alpha$ is -1 for the Nd moment pointing anti-parallel to the magnetic field, and $\alpha=1/3$ for the other three.
The effective Hamiltonian for each Nd is $H_{Nd}^{eff}=-\frac{1}{2}\Delta _{j}^{'}(T,H)\tau _{j}^{z}$, from which it is easy to calculate the average magnetic moment
$\left\langle {{M}_{j}^z} \right\rangle =\tanh \left( \frac{\beta \Delta _{i}^{'}\left( \beta ,H \right)}{2} \right)$.
Now, we can substitute ${{\hat{\tau }}^{z}_j}$ in $H_{fd}$ with $\left\langle {{M}_{j}^z} \right\rangle$.
All the parameters in the model, unless stated otherwise, are given in units of the oxygen mediated nearest neighbor Ir-Ir hopping, $t_{0}$, as commonly used in theories for pyrochlore iridates~\cite{PesinBalents2010,chen2012fd,witczak2013effectiveHamiltonian}.

Figure~\ref{BandStructure}(a) presents the evolution of Ir magnetic moments with field.
While the Nd moments flip from the ``4-0" to the ``3-1" configuration, the Ir moments do not flip with fields up to 7 ~T.
However, their mean field average values are getting smaller, indicating possible canting of the moments from the local z easy-axis.
Their initial configuration depends on the f-d interaction type - ferromagnetic or antiferromagnetic, and determined by the values of $c_1$ and $c_2$.
Nonetheless, both types of interaction yield similar band structures qualitatively.
We present a cut along high symmetry line from the Brillouin zone center $\Gamma=(0,0,0)$ to the edge $L_{1}=\frac{1}{2}(1,1,1)$ in units of reciprocal lattice vectors.
Figures~\ref{BandStructure}(d-g) present the band structure evolution with magnetic field, starting from Weyl semimetal at $H=0$ to Weyl metallic phase at $H>0$ with compensating charge pockets.
While electron pockets are created in the Weyl pair along the $L_{1}$ direction, hole pockets are created in the other three pairs along the high symmetry directions $L_{2}=\frac{1}{2}(1,-1,1)$ and its cyclic variations.

\section{Discussion}
The magnetic field breaks the symmetry between the otherwise equivalent positions in reciprocal space along the line connecting $\Gamma$ and $L$; the field serves as a knob for tuning the valley population.
Figure~\ref{BandStructure}(d) demonstrates two inequivalent pairs of Weyl nodes in the $\Gamma L_{1} L_{2}$ plane.
One pair, hosting electron pockets, is along the $\Gamma L_{1}$ line and the other shifted slightly from the $\Gamma L_{2}$ line hosting hole pockets.
The chemical potential is presented as an orange plane.
Previous study has reported field-induced polarization of Dirac valleys in bismuth~\cite{zhu2012field}, based on the cyclotron motion of the conduction electrons.
In contrast, the valley polarization of Nd$_2$Ir$_2$O$_7$ capitalizes on the magnetic structure of the Ir sublattice in this compound and the amplifying effect of the Nd moments.

\begin{figure}[tbph!]
\includegraphics[width=1\columnwidth]{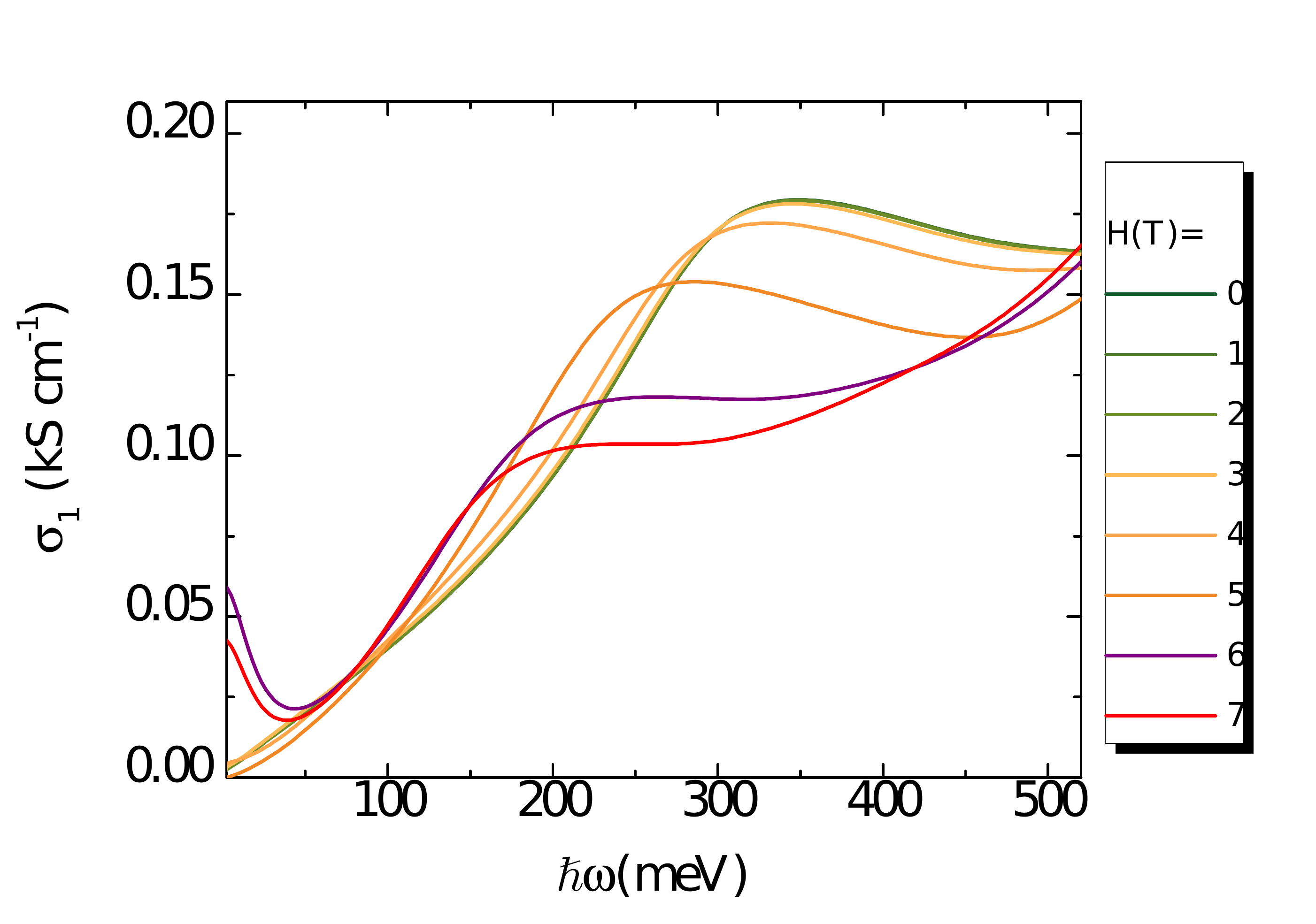}
\caption{Simulated optical conductivity for $\mathbf{U=1.365} $~eV, $\mathbf{c_1=0.026} $~eV and $\mathbf{c_2=-0.039} $~eV. The appearance of Drude spectral weight and an overall red shift of the spectra qualitatively match the experimental data.
In order to match quantitatively, one needs $t_0=0.26 $~eV, suggesting a factor of four mass renormalization of the 5d bands.}
\label{Sigma1Simulation}
\end{figure}

The corresponding optical conductivity for various fields is given in Fig.~\ref{Sigma1Simulation}.
Qualitatively, the theory and the experimental data agree: the simulation predicts the appearance of Drude spectral weight and red shift of the interband transitions upon application of magnetic field.
To match the experimental data one needs $t_{0}=0.26$~eV.
From the dispersion shown in Ref.~\onlinecite{Zhang2017LDA} one can estimate $t_{0}=1.3$~eV,  suggesting a factor of four mass enhancement due to strong correlation effects not captured by the single-particle tight binding Hamiltonian and the Mean Field approximation.
Energy renormalization of this order due to strong correlations is common in transition metal oxides~\cite{basov2011electrodynamics,stricker2014optical,van2008electron}, and observed in SrIrO$_3$~\cite{moon2008SrIrO} and in Sr$_2$IrO$_4$~\cite{wang2018mott}.
Our result is in agreement with angle-resolved photo-emission spectroscopy measurements, that showed bandwidth narrowing with respect to LDA+U calculations.
Nakayama {\it et al.}~\cite{nakayama2016arpes} reported 40~meV width of the occupied band next to the Fermi energy versus about 0.5~eV of the calculated bands~\cite{shinaoka2019DFT}.
Moreover, coupling of the electronic degrees of freedom to the phononic ones would also strongly affect the band structure.
This kind of coupling is clearly evident from our optical conductivity data in the form of asymmetric phonons peaks that also respond asymmetrically to external field.
We mention that similar behavior of band narrowing and electron-boson coupling was observed in Pr$_2$Ir$_2$O$_7$~\cite{kondo2015arpes}.

One of the signatures of the valley imbalance at finite field  (Fig.~\ref{BandStructure}{\bf d} and \ref{BandStructure}{\bf h}) is a Pauli-blocking of the optical conductivity at low frequencies. The fact that the tilted Dirac cones are 3-dimensional makes that the Pauli blocking shows up as a gradual rise starting at $0.05 t_0=13$~meV and ending $0.1 t_0=26$~meV (and an additional one at about half that energy). While such a gradual rise of the optical conductivity could in principle be revealed by the reflectivity, the accuracy obtained in the present study (especially below 30 meV, see Fig.~\ref{Sigma1Data}{\bf c}) does not allow to identify this feature in the experimental data.

Recently some of us reported a $T^2$ free carrier spectral weight as expected for massless Dirac electrons, however the entropy counterpart did not show up in the specific heat~\cite{wang2020unconventional}.
A scenario of temperature dependent charge compensated electron and hole pockets that for $T = 0$ touch at $E_F$, and overlap for finite temperature was dismissed because there are no indications for this in the literature\cite{WanTurner2011,witczak2013effectiveHamiltonian,wang2017weyl}.
However, our present magneto-optics data suggest precisely such a scenario.
In the absence of a magnetic field and at $T = 0$ the Ir 5d bands form Weyl points at $E_F$ and the four Nd moments are in the 4-0 configuration.
At finite temperature the fraction of excited moments is  $n(T)=1/(e^{\Delta_0/T}-1)\approx T/\Delta_0$ where the approximation is valid for $T\gg \Delta_0$.
This creates a random exchange potential $V_{x}$ at the Ir sites, which statistically averages out to zero, but its square has a finite average $V_{x}^2 = J_{fd}^2  T^2 /\Delta_0^2 $ where $J_{fd}$ characterizes the $fd$ exchange interaction.
Substituting $ V_{x}$ for the chemical potential $\mu$  in Eq.
10 of Ashby and Carbotte~\cite{ashby2014chiral}, we obtain for the free carrier spectral weight $\omega_p^2= 2ge^2J_{fd}^2  T^2 /(3\pi\hbar^3\upsilon_F \Delta_0^2) $, where $g$ is the number of Weyl points in the Brillouin zone and $\upsilon_F$ the Fermi velocity.
The only entropy involved is that of the Nd spin degrees of freedom, which is already accounted for by the Schottky anomaly.

\section{Summary and conclusions}
In summary, optical spectroscopy reveals a red shift of the interband transitions and an increase of the Drude spectral weight in a magnetic field below 13~K.
Mean field calculations, taking into account interactions between the 4f and 5d electrons using real measured values extracted from magneto-calorimetry, show that modest magnetic fields are enough to alter the band structure, showing that the rare earth Nds serve as magnetic field ``boosters".
The field creates charge compensated pockets at the four pairs of different Weyl points and induces valley population shift.
While the model of valley imbalance offers a plausible explanation that respects the overall charge neutrality of the system, we do not wish to pretend that it is the only mechanism that could possibly be at play here.
The role of rare earth together with fields of that scale as valley polarizers should be further studied as one of the possible mechanisms for valleytronic devices.
Further experiments such as time domain THz pump probe, Kerr rotation and non-linear electromagnetics~\cite{ma2015mobile} are needed to understand and manipulate the different Weyl points.

\acknowledgments
We thank Nimrod Bachar, Vladimir Kalnizky, Alexey Kuzmenko, Michael Hermele and Gang Chen for fruitful discussions.
This project was supported by the Swiss National Science Foundation through project 200020-179157.
The work at the Institute for Quantum Matter,  an Energy Frontier Research Center was funded by DOE, Office of Science,  Basic Energy Sciences under Award   DE-SC0019331.
This work was partially supported by JST-CREST (JPMJCR18T3).
\appendix

\section{Magneto-reflectance measurements}\label{appendix:magneto-reflectance}

Magneto-reflectance measurements were performed in a cryogenic magnet connected to Fourier transform spectrometer.
Hg and Globar lamps were used as light sources, together with KBr beam splitter.
The iris aperture was 2~mm.
The sample chamber inside the magnet is pumped to high vacuum of about $10^{-8}$~mbar.
The sample and reference mirror were mounted to a sample holder, which in turn was placed on a motorized arm, thus allowing accurate movement between the sample and the mirror alternatively while in the magnet.
Light reflected from the sample or the mirror was collected and measured with a cryogenic Bolometer.
The sample surface is along the $(1,1,1)$ crystallographic direction with area of approximately 1~mm$^2$.
Reflectivity measurements were done in near-to-normal incidence geometry.
The system was cooled down slowly to 4~K, and then warmed up to the desired temperature.
Magnetic field was applied along the $(1,1,1)$ direction from 0~T to 7~T.
In order to prevent misalignment and mechanical movement errors, the sample was measured first with different fields at constant temperature while being kept stable at the same position.
Only then was the mirror measured.
To compensate for drift due to changes with time in the light source and the Bolometer, the reference was measured both before and after the sample had been measured.
The recorded signal, $I(H)$, was used to extract the field dependent reflectivity from the zero field reflectivity via the relation $ R(H)=\frac{I(H)}{I(0)}\cdot R(0)$.
To obtain the dielectric function $\epsilon(\omega)$ and the optical conductivity $\mbox{Re}\sigma(\omega)=(\omega/4\pi)\mbox{Im}\epsilon(\omega)$ we fitted the Drude-Lorentz expansion
\begin{equation}
\epsilon(\omega)=\epsilon_{\infty} + \sum_{j}\frac{\omega_{p,j}^2}{\omega_{0,j}^2-\omega^2-i\gamma_{j}\omega}
\label{epsilon}
\nonumber
\end{equation}
to the magneto-reflectivity (15-90~meV), the zero field reflectivity (90-500~meV), the zero field room temperature ellipsometery (0.5-2.5~eV) and the DC conductivity.
For the Drude term $\omega_{0}=0$.

\section{Specific heat}\label{appendix:cv}
Specific heat was measured using a PPMS$^{\copyright}$.
First, the thermometers were calibrated under all the desired magnetic fields to correctly read the sample temperature.
Next, the heat capacity of the addenda, $C_{add}$, was measured for each field, including a small amount of vacuum grease used as glue and thermal link between the platform and the sample.
Finally, the sample including the addenda were measured yielding $C_{tot}$ from which the heat capacity of the sample alone is calculated via $C_{sample}(H) = C_{tot}(H) - C_{add}(H)$.
These values were converted to units of JK$^{-2}$mol$^{-1}$ using the sample mass and the compound molar mass.

\begin{figure}[!!t]
\includegraphics[width=\columnwidth]{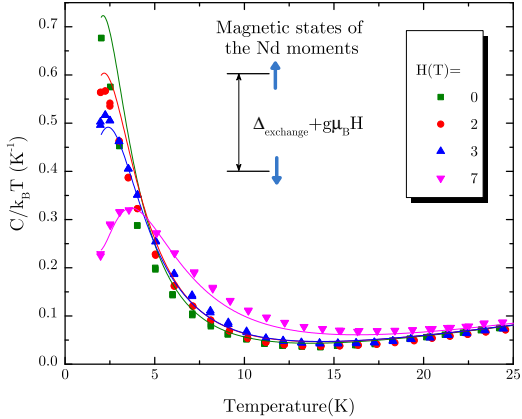}
\caption{Specific heat data as function of magnetic field.
The lines are the fits to data with single type 1 domain.\label{SpecificHeatData}}
\end{figure} 
The exchange potential was calculated from the specific heat data.
The Nd ions occupy the corners of a tetrahedron.
We use the label $j=0$ for the Nd atom on the $(1,1,1)$ corner, and $j=1,2,3$ for the other three.
The relevant states for the specific heat form spin-orbital doublet polarized along the local quantization axis, which for $j=0$ is $(1,1,1)$.
Magnetic order of the Ir sublattice induces an exchange field at the Nd sites oriented along the local quantization axis.
Two types of domains are possible, which for a given tetrahedron correspond to exchange fields either pointing all in (type 1 domain) or all out (type 2 domain).
At low field these domains alternate and are separated by domain walls.
Here we found that a field of 3 tesla suffices to favor one type of domain and suppress all domain walls.

An external magnetic field $H$ along $(1,1,1)$ has a projection $f_jH$ on the local quantization axis, where $f_0=1$, and for $j=1,2,3$ $f_j=-1/3$.
Together with the local exchange field this gives the energy splitting of the $4f$ doublet
\begin{equation}
\Delta_{j}= \sigma V_{df} +g f_j \mu_BH 
\nonumber
\end{equation}
where $g=2.4$ is the gyromagnetic ratio of Nd$^{3+}$, and $\sigma = 1  (-1)$ corresponds to domains of type 1 (2).
 
The specific heat is 
\begin{equation}
\frac{C(T)}{k_B T}=\frac{d}{dT}
\sum_{j=0}^3 \sum_{s=\pm 1}\frac{\ln{\left(1+e^{s\Delta_{j}/T}\right)}}{1+e^{s\Delta_{j}/T}}
+
aT^2 - bT^4
\label{eq:cv}
\nonumber
\end{equation}
where the last two terms describe the phonon contribution.
We fitted this expression to the experimental specific heat data by adjusting $a$, $b$, and $V_{df}$.

\begin{figure}[!!t]
\includegraphics[width=\columnwidth]{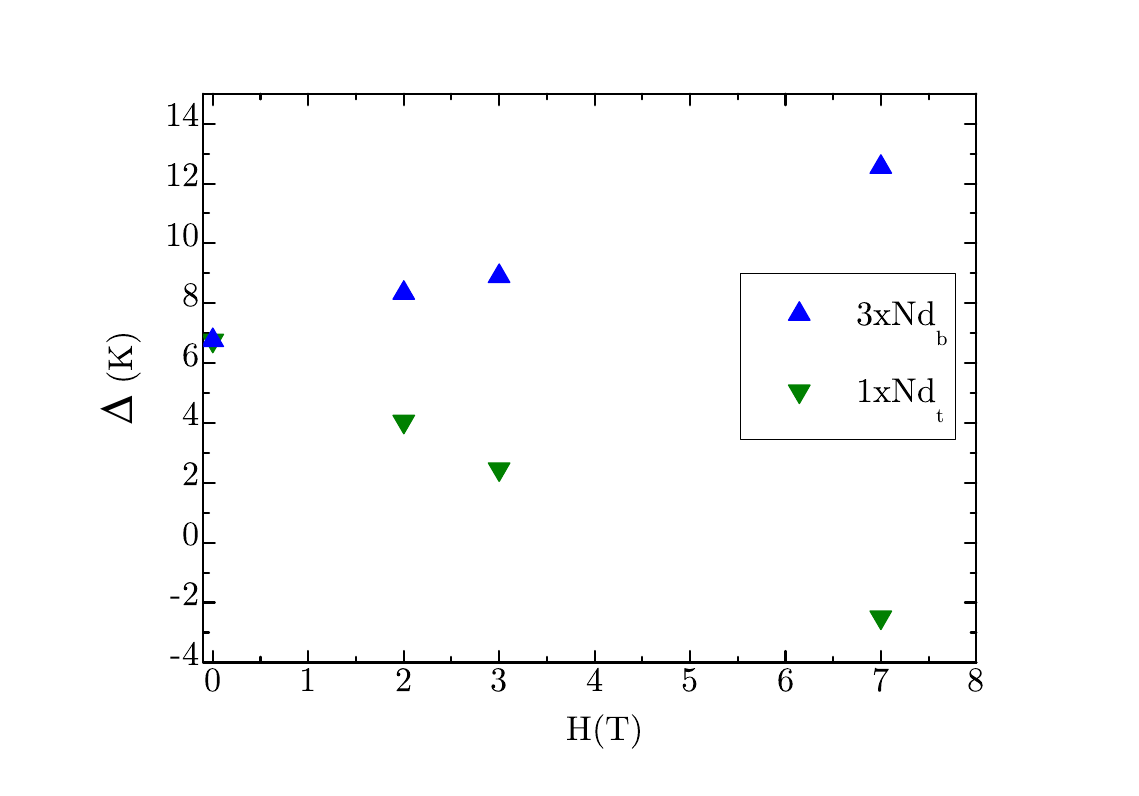}
\caption{Exchange potential at the Nd sites.
\label{SpecificHeatFit}}
\end{figure}

In Fig.~\ref{SpecificHeatData} $C(T)/T$ is presented for different magnetic fields together with the fits, assuming $\sigma=1$ (type 1 domains).
The corresponding  $\Delta_{0}$ and $\Delta_{1}$ is shown in Fig.~\ref{SpecificHeatFit}.
The field dependent exchange coupling is given by $V_{df} = 6.7 + 0.285H$ in units of kelvin and with $H$ in units of tesla.

Fitting the data with $\sigma=-1$ (type 2 domains) or an equal mixture of both types of domains,  gives unreasonably high values of $V_{df}$.
With these parameters we calculate the thermally averaged spin values of the Nd atoms
\begin{equation}
{M}_{j}^z= \arctan\left( \frac{\Delta_{j}}{2T} \right)
\nonumber
\end{equation}
which are inserted in Eq.~1 of the main text to calculate the bandstructure in the presence the Nd exchange potential on the Ir sites.

\section{The electron band structure}\label{appendix:bands}
The electron band structure was calculated using the tight binding Hamiltonian of Ref.~19 of the main text, which describes electrons with pseudospin $j=1/2$ having an onsite Hubbard U interaction, $H_{U}$, and the Zeeman term:
\begin{align}
H_0  = \sum\limits_{s,s'} \sum\limits_{\mathbf{R}}\sum\limits_{\left\langle i,j \right\rangle} d_{\mathbf{R}is}^{\dagger}\left(t_1\mathbf{I}+it_2 \mathbf{d}_{ij} \cdot\pmb{\sigma}\right)_{ss'} d_{\mathbf{R}js'} + \nonumber \\
+ \sum\limits_{\left\langle \left\langle i,j \right\rangle\right\rangle} d_{\mathbf{R}is}^{\dagger} \left(t_{1}^{'}\mathbf{I}+i\left[ t_{2}^{'}{{\mathcal{R}}_{ij}}+t_{3}^{'}{{\mathcal{D}}_{ij}} \right]\cdot \pmb{\sigma} \right)_{ss'} d_{\mathbf{R}js'} .
\end{align}
Here $d_{\mathbf{R}is}^{\dagger}$ creates an electron at unit cell $\mathbf{R}$, at Ir site $i$ with pseudospin $s$, $t_1, t_2$ and $t_{1}^{'}, t_{2}^{'}, t_{3}^{'}$ are the NN and NNN hopping amplitudes, respectively, $\pmb{\sigma}$ is a vector of the Pauli matrices and $\mathbf{d}_{ij}, {{\mathcal{R}}_{ij}}, {{\mathcal{D}}_{ij}}$ are real geometrical vectors.
Mean field decoupling of the Hubbard term gives
\begin{align}
{{H}_{U}}=\sum\limits_{\mathbf{R},i}
\left[
\frac{2U}{3}{{\left\langle {{\mathbf{j}}_{\mathbf{R},i}} \right\rangle }^{2}}-\frac{4U}{3}\left\langle {{\mathbf{j}}_{\mathbf{R},i}} \right\rangle \cdot {{\mathbf{j}}_{\mathbf{R},i}}
\right],
\nonumber
\end{align} 
with ${{\mathbf{j}}_{\mathbf{R},i}}=\frac{1}{2}\sum\limits_{\alpha ,\beta \in \left\{ \uparrow ,\downarrow  \right\}}{d_{\mathbf{R},i,\alpha }^{\dagger }{{\mathbf{\sigma }}_{\alpha \beta }}{{d}_{\mathbf{R},i,\beta }}}$ the pseudospin operator.
We add the f-d interaction term, as described in the main text, and solve with four mean field parameters for the Ir sublattice pseudospins.
The complex optical conductivity is calculated as follows:
\begin{eqnarray}
\sigma (\omega )&=&
\frac{q_e^2}{\hbar \Omega}
\sum\limits_{\mathbf{k},j}^{1^{st}BZ}
\mathbf{v}_{j,j}(\mathbf{k})
\cdot 
\mathbf{v}_{j,j}(\mathbf{k})
\left( 
-\frac{\partial f_{\mathbf{k},j}}{\partial \varepsilon_{\mathbf{k},j}} 
\right)
\frac{i}{\omega +i\delta}
\nonumber\\
&+&
\frac{q_e^2} {\hbar \Omega}  
\sum\limits_{\mathbf{k},j\ne m}^{{1^{st}}BZ}
\mathbf{v}_{j,m}(\mathbf{k})\cdot \mathbf{v}_{m,j}(\mathbf{k}) 
\left( \frac{f_{\mathbf{k},j}-f_{\mathbf{k},m}}{\varepsilon_{\mathbf{k},m} - \varepsilon_{\mathbf{k},j}} \right)
\nonumber\\
&\times&
\frac{i\omega }{\omega \left( \omega +i\delta  \right)-
\left(\varepsilon_{\mathbf{k},m} - \varepsilon_{\mathbf{k},j} \right)^2}. 
\end{eqnarray}

\begin{align*}
&\mbox{where } 
\hbar {{\mathbf{v}}_{j,m}}(\mathbf{k})= \\  &\sum\limits_{\eta ,\mu }^{{}}{u_{\eta ,j}^{*}(\mathbf{h}){{u}_{\mu ,m}}(\mathbf{k})\left[ \frac{\partial {{H}_{\eta ,\mu }}(\mathbf{k})}{\partial \mathbf{k}}+i\left( {{\mathbf{d}}_{\mu }}-{{\mathbf{d}}_{\eta }} \right){{H}_{\eta ,\mu }}(\mathbf{k}) \right]},
\end{align*}
and $\mathbf{d}_{\mu}$ are the positions of the four Ir atoms in a unit cell and ${u}_{\mu ,m}$ are the matrix elements of the unitary diagonalizing matrix of the Hamiltonian.
%
%

\begin{thebibliography}{33}%
\makeatletter
\providecommand \@ifxundefined [1]{%
 \@ifx{#1\undefined}
}%
\providecommand \@ifnum [1]{%
 \ifnum #1\expandafter \@firstoftwo
 \else \expandafter \@secondoftwo
 \fi
}%
\providecommand \@ifx [1]{%
 \ifx #1\expandafter \@firstoftwo
 \else \expandafter \@secondoftwo
 \fi
}%
\providecommand \natexlab [1]{#1}%
\providecommand \enquote  [1]{``#1''}%
\providecommand \bibnamefont  [1]{#1}%
\providecommand \bibfnamefont [1]{#1}%
\providecommand \citenamefont [1]{#1}%
\providecommand \href@noop [0]{\@secondoftwo}%
\providecommand \href [0]{\begingroup \@sanitize@url \@href}%
\providecommand \@href[1]{\@@startlink{#1}\@@href}%
\providecommand \@@href[1]{\endgroup#1\@@endlink}%
\providecommand \@sanitize@url [0]{\catcode `\\12\catcode `\$12\catcode
  `\&12\catcode `\#12\catcode `\^12\catcode `\_12\catcode `\%12\relax}%
\providecommand \@@startlink[1]{}%
\providecommand \@@endlink[0]{}%
\providecommand \url  [0]{\begingroup\@sanitize@url \@url }%
\providecommand \@url [1]{\endgroup\@href {#1}{\urlprefix }}%
\providecommand \urlprefix  [0]{URL }%
\providecommand \Eprint [0]{\href }%
\providecommand \doibase [0]{http://dx.doi.org/}%
\providecommand \selectlanguage [0]{\@gobble}%
\providecommand \bibinfo  [0]{\@secondoftwo}%
\providecommand \bibfield  [0]{\@secondoftwo}%
\providecommand \translation [1]{[#1]}%
\providecommand \BibitemOpen [0]{}%
\providecommand \bibitemStop [0]{}%
\providecommand \bibitemNoStop [0]{.\EOS\space}%
\providecommand \EOS [0]{\spacefactor3000\relax}%
\providecommand \BibitemShut  [1]{\csname bibitem#1\endcsname}%
\let\auto@bib@innerbib\@empty
\bibitem [{\citenamefont {Vafek}\ and\ \citenamefont
  {Vishwanath}(2014)}]{vafek2014dirac}%
  \BibitemOpen
  \bibfield  {author} {\bibinfo {author} {\bibfnamefont {Oskar}\ \bibnamefont
  {Vafek}}\ and\ \bibinfo {author} {\bibfnamefont {Ashvin}\ \bibnamefont
  {Vishwanath}},\ }\bibfield  {title} {\enquote {\bibinfo {title} {{Dirac
  Fermions in Solids: From High-Tc Cuprates and Graphene to Topological
  Insulators and Weyl Semimetals}},}\ }\href
  {https://doi.org/10.1146/annurev-conmatphys-031113-133841} {\bibfield
  {journal} {\bibinfo  {journal} {Annual Review of Condensed Matter Physics}\
  }\textbf {\bibinfo {volume} {5}},\ \bibinfo {pages} {83--112} (\bibinfo
  {year} {2014})}\BibitemShut {NoStop}%
\bibitem [{\citenamefont {Armitage}\ \emph {et~al.}(2018)\citenamefont
  {Armitage}, \citenamefont {Mele},\ and\ \citenamefont
  {Vishwanath}}]{armitage2018weyl}%
  \BibitemOpen
  \bibfield  {author} {\bibinfo {author} {\bibfnamefont {N.~P.}\ \bibnamefont
  {Armitage}}, \bibinfo {author} {\bibfnamefont {E.~J.}\ \bibnamefont {Mele}},
  \ and\ \bibinfo {author} {\bibfnamefont {Ashvin}\ \bibnamefont
  {Vishwanath}},\ }\bibfield  {title} {\enquote {\bibinfo {title} {{Weyl and
  Dirac semimetals in three-dimensional solids}},}\ }\href
  {https://doi.org/10.1103/RevModPhys.90.015001} {\bibfield  {journal}
  {\bibinfo  {journal} {Reviews of Modern Physics}\ }\textbf {\bibinfo {volume}
  {90}},\ \bibinfo {pages} {015001} (\bibinfo {year} {2018})}\BibitemShut
  {NoStop}%
\bibitem [{\citenamefont {Lv}\ \emph {et~al.}(2015)\citenamefont {Lv},
  \citenamefont {Weng}, \citenamefont {Fu}, \citenamefont {Wang}, \citenamefont
  {Miao}, \citenamefont {Ma}, \citenamefont {Richard}, \citenamefont {Huang},
  \citenamefont {Zhao}, \citenamefont {Chen}, \citenamefont {Fang},
  \citenamefont {Dai}, \citenamefont {Qian},\ and\ \citenamefont
  {Ding}}]{lv2015experimental}%
  \BibitemOpen
  \bibfield  {author} {\bibinfo {author} {\bibfnamefont {B.~Q.}\ \bibnamefont
  {Lv}}, \bibinfo {author} {\bibfnamefont {H.~M.}\ \bibnamefont {Weng}},
  \bibinfo {author} {\bibfnamefont {B.~B.}\ \bibnamefont {Fu}}, \bibinfo
  {author} {\bibfnamefont {X.~P.}\ \bibnamefont {Wang}}, \bibinfo {author}
  {\bibfnamefont {H.}~\bibnamefont {Miao}}, \bibinfo {author} {\bibfnamefont
  {J.}~\bibnamefont {Ma}}, \bibinfo {author} {\bibfnamefont {P.}~\bibnamefont
  {Richard}}, \bibinfo {author} {\bibfnamefont {X.~C.}\ \bibnamefont {Huang}},
  \bibinfo {author} {\bibfnamefont {L.~X.}\ \bibnamefont {Zhao}}, \bibinfo
  {author} {\bibfnamefont {G.~F.}\ \bibnamefont {Chen}}, \bibinfo {author}
  {\bibfnamefont {Z.}~\bibnamefont {Fang}}, \bibinfo {author} {\bibfnamefont
  {X.}~\bibnamefont {Dai}}, \bibinfo {author} {\bibfnamefont {T.}~\bibnamefont
  {Qian}}, \ and\ \bibinfo {author} {\bibfnamefont {H.}~\bibnamefont {Ding}},\
  }\bibfield  {title} {\enquote {\bibinfo {title} {{Experimental discovery of
  Weyl semimetal TaAs}},}\ }\href {https://doi.org/10.1103/PhysRevX.5.031013}
  {\bibfield  {journal} {\bibinfo  {journal} {Physical Review X}\ }\textbf
  {\bibinfo {volume} {5}},\ \bibinfo {pages} {031013} (\bibinfo {year}
  {2015})}\BibitemShut {NoStop}%
\bibitem [{\citenamefont {Nakatsuji}\ \emph {et~al.}(2015)\citenamefont
  {Nakatsuji}, \citenamefont {Kiyohara},\ and\ \citenamefont
  {Higo}}]{nakatsuji2015large}%
  \BibitemOpen
  \bibfield  {author} {\bibinfo {author} {\bibfnamefont {Satoru}\ \bibnamefont
  {Nakatsuji}}, \bibinfo {author} {\bibfnamefont {Naoki}\ \bibnamefont
  {Kiyohara}}, \ and\ \bibinfo {author} {\bibfnamefont {Tomoya}\ \bibnamefont
  {Higo}},\ }\bibfield  {title} {\enquote {\bibinfo {title} {{Large anomalous
  Hall effect in a non-collinear antiferromagnet at room temperature}},}\
  }\href {https://doi.org/10.1038/nature15723} {\bibfield  {journal} {\bibinfo
  {journal} {Nature}\ }\textbf {\bibinfo {volume} {527}},\ \bibinfo {pages}
  {212--215} (\bibinfo {year} {2015})}\BibitemShut {NoStop}%
\bibitem [{\citenamefont {Rao}\ \emph {et~al.}(2019)\citenamefont {Rao},
  \citenamefont {Li}, \citenamefont {Zhang}, \citenamefont {Tian},
  \citenamefont {Li}, \citenamefont {Fu}, \citenamefont {Tang}, \citenamefont
  {Wang}, \citenamefont {Li}, \citenamefont {Fan}, \citenamefont {Li},
  \citenamefont {Huang}, \citenamefont {Liu}, \citenamefont {Long},
  \citenamefont {Weng}, \citenamefont {Shi}, \citenamefont {Lei}, \citenamefont
  {Sun}, \citenamefont {Qian},\ and\ \citenamefont
  {Ding}}]{rao2019observation}%
  \BibitemOpen
  \bibfield  {author} {\bibinfo {author} {\bibfnamefont {Zhicheng}\
  \bibnamefont {Rao}}, \bibinfo {author} {\bibfnamefont {Hang}\ \bibnamefont
  {Li}}, \bibinfo {author} {\bibfnamefont {Tiantian}\ \bibnamefont {Zhang}},
  \bibinfo {author} {\bibfnamefont {Shangjie}\ \bibnamefont {Tian}}, \bibinfo
  {author} {\bibfnamefont {Chenghe}\ \bibnamefont {Li}}, \bibinfo {author}
  {\bibfnamefont {Binbin}\ \bibnamefont {Fu}}, \bibinfo {author} {\bibfnamefont
  {Cenyao}\ \bibnamefont {Tang}}, \bibinfo {author} {\bibfnamefont
  {Le}~\bibnamefont {Wang}}, \bibinfo {author} {\bibfnamefont {Zhilin}\
  \bibnamefont {Li}}, \bibinfo {author} {\bibfnamefont {Wenhui}\ \bibnamefont
  {Fan}}, \bibinfo {author} {\bibfnamefont {Jiajun}\ \bibnamefont {Li}},
  \bibinfo {author} {\bibfnamefont {Yaobo}\ \bibnamefont {Huang}}, \bibinfo
  {author} {\bibfnamefont {Zhehong}\ \bibnamefont {Liu}}, \bibinfo {author}
  {\bibfnamefont {Chen}\ \bibnamefont {Long}, \bibfnamefont {Youwen
  and ~Fang}}, \bibinfo {author} {\bibfnamefont {Hongming}\ \bibnamefont
  {Weng}}, \bibinfo {author} {\bibfnamefont {Youguo}\ \bibnamefont {Shi}},
  \bibinfo {author} {\bibfnamefont {Hechang}\ \bibnamefont {Lei}}, \bibinfo
  {author} {\bibfnamefont {Yujie}\ \bibnamefont {Sun}}, \bibinfo {author}
  {\bibfnamefont {Tian}\ \bibnamefont {Qian}}, \ and\ \bibinfo {author}
  {\bibfnamefont {Hong}\ \bibnamefont {Ding}},\ }\bibfield  {title} {\enquote
  {\bibinfo {title} {{Observation of unconventional chiral fermions with long
  Fermi arcs in CoSi}},}\ }\href {https://doi.org/10.1038/s41586-019-1031-8}
  {\bibfield  {journal} {\bibinfo  {journal} {Nature}\ }\textbf {\bibinfo
  {volume} {567}},\ \bibinfo {pages} {496--499} (\bibinfo {year}
  {2019})}\BibitemShut {NoStop}%
\bibitem [{\citenamefont {Richard}\ \emph {et~al.}(2010)\citenamefont
  {Richard}, \citenamefont {Nakayama}, \citenamefont {Sato}, \citenamefont
  {Neupane}, \citenamefont {Xu}, \citenamefont {Bowen}, \citenamefont {Chen},
  \citenamefont {Luo}, \citenamefont {Wang}, \citenamefont {Dai}, \citenamefont
  {Fang}, \citenamefont {Ding},\ and\ \citenamefont
  {Takahashi}}]{richard2010observation}%
  \BibitemOpen
  \bibfield  {author} {\bibinfo {author} {\bibfnamefont {P.}~\bibnamefont
  {Richard}}, \bibinfo {author} {\bibfnamefont {K.}~\bibnamefont {Nakayama}},
  \bibinfo {author} {\bibfnamefont {T.}~\bibnamefont {Sato}}, \bibinfo {author}
  {\bibfnamefont {M.}~\bibnamefont {Neupane}}, \bibinfo {author} {\bibfnamefont
  {Y.-M.}\ \bibnamefont {Xu}}, \bibinfo {author} {\bibfnamefont {J.~H.}\
  \bibnamefont {Bowen}}, \bibinfo {author} {\bibfnamefont {G.~F.}\ \bibnamefont
  {Chen}}, \bibinfo {author} {\bibfnamefont {J.~L.}\ \bibnamefont {Luo}},
  \bibinfo {author} {\bibfnamefont {N.~L.}\ \bibnamefont {Wang}}, \bibinfo
  {author} {\bibfnamefont {X.}~\bibnamefont {Dai}}, \bibinfo {author}
  {\bibfnamefont {Z.}~\bibnamefont {Fang}}, \bibinfo {author} {\bibfnamefont
  {H.}~\bibnamefont {Ding}}, \ and\ \bibinfo {author} {\bibfnamefont
  {T.}~\bibnamefont {Takahashi}},\ }\bibfield  {title} {\enquote {\bibinfo
  {title} {{Observation of Dirac cone electronic dispersion in
  BaFe$_2$As$_2$}},}\ }\href {https://doi.org/10.1103/PhysRevLett.104.137001}
  {\bibfield  {journal} {\bibinfo  {journal} {Phys. Rev. Lett.}\ }\textbf
  {\bibinfo {volume} {104}},\ \bibinfo {pages} {137001} (\bibinfo {year}
  {2010})}\BibitemShut {NoStop}%
\bibitem [{\citenamefont {Borne}\ \emph {et~al.}(2010)\citenamefont {Borne},
  \citenamefont {Carbotte},\ and\ \citenamefont {Nicol}}]{borne2010specific}%
  \BibitemOpen
  \bibfield  {author} {\bibinfo {author} {\bibfnamefont {A.~J.~H.}\
  \bibnamefont {Borne}}, \bibinfo {author} {\bibfnamefont {J.~P.}\ \bibnamefont
  {Carbotte}}, \ and\ \bibinfo {author} {\bibfnamefont {E.~J.}\ \bibnamefont
  {Nicol}},\ }\bibfield  {title} {\enquote {\bibinfo {title} {{Specific heat
  across the superconducting dome in the cuprates}},}\ }\href
  {https://doi.org/10.1103/PhysRevB.82.094523} {\bibfield  {journal} {\bibinfo
  {journal} {Phys. Rev. B}\ }\textbf {\bibinfo {volume} {82}},\ \bibinfo
  {pages} {094523} (\bibinfo {year} {2010})}\BibitemShut {NoStop}%
\bibitem [{\citenamefont {Vitale}\ \emph {et~al.}(2018)\citenamefont {Vitale},
  \citenamefont {Nezich}, \citenamefont {Varghese}, \citenamefont {Kim},
  \citenamefont {Gedik}, \citenamefont {Jarillo-Herrero}, \citenamefont
  {Xiao},\ and\ \citenamefont {Rothschild}}]{vitale2018valleytronics}%
  \BibitemOpen
  \bibfield  {author} {\bibinfo {author} {\bibfnamefont {Steven~A}\
  \bibnamefont {Vitale}}, \bibinfo {author} {\bibfnamefont {Daniel}\
  \bibnamefont {Nezich}}, \bibinfo {author} {\bibfnamefont {Joseph~O}\
  \bibnamefont {Varghese}}, \bibinfo {author} {\bibfnamefont {Philip}\
  \bibnamefont {Kim}}, \bibinfo {author} {\bibfnamefont {Nuh}\ \bibnamefont
  {Gedik}}, \bibinfo {author} {\bibfnamefont {Pablo}\ \bibnamefont
  {Jarillo-Herrero}}, \bibinfo {author} {\bibfnamefont {Di}~\bibnamefont
  {Xiao}}, \ and\ \bibinfo {author} {\bibfnamefont {Mordechai}\ \bibnamefont
  {Rothschild}},\ }\bibfield  {title} {\enquote {\bibinfo {title}
  {{Valleytronics: opportunities, challenges, and paths forward}},}\ }\href
  {https://doi.org/10.1002/smll.201801483} {\bibfield  {journal} {\bibinfo
  {journal} {Small}\ }\textbf {\bibinfo {volume} {14}},\ \bibinfo {pages}
  {1801483} (\bibinfo {year} {2018})}\BibitemShut {NoStop}%
\bibitem [{\citenamefont {Wan}\ \emph {et~al.}(2011)\citenamefont {Wan},
  \citenamefont {Turner}, \citenamefont {Vishwanath},\ and\ \citenamefont
  {Savrasov}}]{WanTurner2011}%
  \BibitemOpen
  \bibfield  {author} {\bibinfo {author} {\bibfnamefont {Xiangang}\
  \bibnamefont {Wan}}, \bibinfo {author} {\bibfnamefont {Ari~M.}\ \bibnamefont
  {Turner}}, \bibinfo {author} {\bibfnamefont {Ashvin}\ \bibnamefont
  {Vishwanath}}, \ and\ \bibinfo {author} {\bibfnamefont {Sergey~Y.}\
  \bibnamefont {Savrasov}},\ }\bibfield  {title} {\enquote {\bibinfo {title}
  {{Topological semimetal and Fermi-arc surface states in the electronic
  structure of pyrochlore iridates}},}\ }\href
  {https://doi.org/10.1103/PhysRevB.83.205101} {\bibfield  {journal} {\bibinfo
  {journal} {Phys. Rev. B}\ }\textbf {\bibinfo {volume} {83}},\ \bibinfo
  {pages} {205101} (\bibinfo {year} {2011})}\BibitemShut {NoStop}%
\bibitem [{\citenamefont {Li}\ \emph {et~al.}(2021)\citenamefont {Li},
  \citenamefont {Oh}, \citenamefont {Son}, \citenamefont {Song}, \citenamefont
  {Kim}, \citenamefont {Song}, \citenamefont {Kim}, \citenamefont {Chang},
  \citenamefont {Kim}, \citenamefont {Yang},\ and\ \citenamefont
  {Noh}}]{li2021correlated}%
  \BibitemOpen
  \bibfield  {author} {\bibinfo {author} {\bibfnamefont {Yangyang}\
  \bibnamefont {Li}}, \bibinfo {author} {\bibfnamefont {Taekoo}\ \bibnamefont
  {Oh}}, \bibinfo {author} {\bibfnamefont {Jaeseok}\ \bibnamefont {Son}},
  \bibinfo {author} {\bibfnamefont {Jeongkeun}\ \bibnamefont {Song}}, \bibinfo
  {author} {\bibfnamefont {Mi~Kyung}\ \bibnamefont {Kim}}, \bibinfo {author}
  {\bibfnamefont {Dongjun}\ \bibnamefont {Song}}, \bibinfo {author}
  {\bibfnamefont {Sukhyun}\ \bibnamefont {Kim}}, \bibinfo {author}
  {\bibfnamefont {Seo~Hyoung}\ \bibnamefont {Chang}}, \bibinfo {author}
  {\bibfnamefont {Changyoung}\ \bibnamefont {Kim}}, \bibinfo {author}
  {\bibfnamefont {Bohm-Jung}\ \bibnamefont {Yang}}, \ and\ \bibinfo {author}
  {\bibfnamefont {Tae~Won}\ \bibnamefont {Noh}},\ }\bibfield  {title} {\enquote
  {\bibinfo {title} {{Correlated Magnetic Weyl Semimetal State in Strained
  Pr$_2$Ir$_2$O$_7$}},}\ }\href {https://doi.org/10.1002/adma.202008528}
  {\bibfield  {journal} {\bibinfo  {journal} {Advanced Materials}\ ,\ \bibinfo
  {pages} {2008528}} (\bibinfo {year} {2021})}\BibitemShut {NoStop}%
\bibitem [{\citenamefont {Ohtsuki}\ \emph {et~al.}(2019)\citenamefont
  {Ohtsuki}, \citenamefont {Tian}, \citenamefont {Endo}, \citenamefont {Halim},
  \citenamefont {Katsumoto}, \citenamefont {Kohama}, \citenamefont {Kindo},
  \citenamefont {Lippmaa},\ and\ \citenamefont
  {Nakatsuji}}]{ohtsuki2019strain}%
  \BibitemOpen
  \bibfield  {author} {\bibinfo {author} {\bibfnamefont {Takumi}\ \bibnamefont
  {Ohtsuki}}, \bibinfo {author} {\bibfnamefont {Zhaoming}\ \bibnamefont
  {Tian}}, \bibinfo {author} {\bibfnamefont {Akira}\ \bibnamefont {Endo}},
  \bibinfo {author} {\bibfnamefont {Mario}\ \bibnamefont {Halim}}, \bibinfo
  {author} {\bibfnamefont {Shingo}\ \bibnamefont {Katsumoto}}, \bibinfo
  {author} {\bibfnamefont {Yoshimitsu}\ \bibnamefont {Kohama}}, \bibinfo
  {author} {\bibfnamefont {Koichi}\ \bibnamefont {Kindo}}, \bibinfo {author}
  {\bibfnamefont {Mikk}\ \bibnamefont {Lippmaa}}, \ and\ \bibinfo {author}
  {\bibfnamefont {Satoru}\ \bibnamefont {Nakatsuji}},\ }\bibfield  {title}
  {\enquote {\bibinfo {title} {{Strain-induced spontaneous Hall effect in an
  epitaxial thin film of a Luttinger semimetal}},}\ }\href
  {https://doi.org/10.1073/pnas.1819489116} {\bibfield  {journal} {\bibinfo
  {journal} {Proceedings of the National Academy of Sciences}\ }\textbf
  {\bibinfo {volume} {116}},\ \bibinfo {pages} {8803--8808} (\bibinfo {year}
  {2019})}\BibitemShut {NoStop}%
\bibitem [{\citenamefont {Tian}\ \emph {et~al.}(2016)\citenamefont {Tian},
  \citenamefont {Kohama}, \citenamefont {Tomita}, \citenamefont {Ishizuka},
  \citenamefont {Hsieh}, \citenamefont {Ishikawa}, \citenamefont {Kindo},
  \citenamefont {Balents},\ and\ \citenamefont {Nakatsuji}}]{tian2016field}%
  \BibitemOpen
  \bibfield  {author} {\bibinfo {author} {\bibfnamefont {Zhaoming}\
  \bibnamefont {Tian}}, \bibinfo {author} {\bibfnamefont {Yoshimitsu}\
  \bibnamefont {Kohama}}, \bibinfo {author} {\bibfnamefont {Takahiro}\
  \bibnamefont {Tomita}}, \bibinfo {author} {\bibfnamefont {Hiroaki}\
  \bibnamefont {Ishizuka}}, \bibinfo {author} {\bibfnamefont {Timothy~H}\
  \bibnamefont {Hsieh}}, \bibinfo {author} {\bibfnamefont {Jun~J}\ \bibnamefont
  {Ishikawa}}, \bibinfo {author} {\bibfnamefont {Koichi}\ \bibnamefont
  {Kindo}}, \bibinfo {author} {\bibfnamefont {Leon}\ \bibnamefont {Balents}}, \
  and\ \bibinfo {author} {\bibfnamefont {Satoru}\ \bibnamefont {Nakatsuji}},\
  }\bibfield  {title} {\enquote {\bibinfo {title} {{Field-induced quantum
  metal--insulator transition in the pyrochlore iridate Nd$_2$Ir$_2$O$_7$}},}\
  }\href {https://doi.org/10.1038/nphys3567} {\bibfield  {journal} {\bibinfo
  {journal} {Nature Physics}\ }\textbf {\bibinfo {volume} {12}},\ \bibinfo
  {pages} {134--138} (\bibinfo {year} {2016})}\BibitemShut {NoStop}%
\bibitem [{\citenamefont {Wang}\ \emph {et~al.}(2020)\citenamefont {Wang},
  \citenamefont {Xu}, \citenamefont {Rischau}, \citenamefont {Bachar},
  \citenamefont {Michon}, \citenamefont {Teyssier}, \citenamefont {Qiu},
  \citenamefont {Ohtsuki}, \citenamefont {Cheng}, \citenamefont {Armitage},
  \citenamefont {Nakatsuji},\ and\ \citenamefont {van~der
  Marel}}]{wang2020unconventional}%
  \BibitemOpen
  \bibfield  {author} {\bibinfo {author} {\bibfnamefont {K.}~\bibnamefont
  {Wang}}, \bibinfo {author} {\bibfnamefont {Bing}\ \bibnamefont {Xu}},
  \bibinfo {author} {\bibfnamefont {C.~W.}\ \bibnamefont {Rischau}}, \bibinfo
  {author} {\bibfnamefont {N.}~\bibnamefont {Bachar}}, \bibinfo {author}
  {\bibfnamefont {B.}~\bibnamefont {Michon}}, \bibinfo {author} {\bibfnamefont
  {J.}~\bibnamefont {Teyssier}}, \bibinfo {author} {\bibfnamefont
  {Y.}~\bibnamefont {Qiu}}, \bibinfo {author} {\bibfnamefont {T.}~\bibnamefont
  {Ohtsuki}}, \bibinfo {author} {\bibfnamefont {Bing}\ \bibnamefont {Cheng}},
  \bibinfo {author} {\bibfnamefont {N.~P.}\ \bibnamefont {Armitage}}, \bibinfo
  {author} {\bibfnamefont {S.}~\bibnamefont {Nakatsuji}}, \ and\ \bibinfo
  {author} {\bibfnamefont { ~D.}\ \bibnamefont {van~der Marel}},\ }\bibfield
  {title} {\enquote {\bibinfo {title} {{Unconventional free charge in the
  correlated semimetal Nd$_2$Ir$_2$O$_7$}},}\ }\href
  {https://doi.org/10.1038/s41567-020-0955-0} {\bibfield  {journal} {\bibinfo
  {journal} {Nature Physics}\ ,\ \bibinfo {pages} {1--5}} (\bibinfo {year}
  {2020})}\BibitemShut {NoStop}%
\bibitem [{\citenamefont {Ma}\ \emph {et~al.}(2015)\citenamefont {Ma},
  \citenamefont {Cui}, \citenamefont {Ueda}, \citenamefont {Tang},
  \citenamefont {Chen}, \citenamefont {Tamura}, \citenamefont {Wu},
  \citenamefont {Fujioka}, \citenamefont {Tokura},\ and\ \citenamefont
  {Shen}}]{ma2015mobile}%
  \BibitemOpen
  \bibfield  {author} {\bibinfo {author} {\bibfnamefont {Eric~Yue}\
  \bibnamefont {Ma}}, \bibinfo {author} {\bibfnamefont {Yong-Tao}\ \bibnamefont
  {Cui}}, \bibinfo {author} {\bibfnamefont {Kentaro}\ \bibnamefont {Ueda}},
  \bibinfo {author} {\bibfnamefont {Shujie}\ \bibnamefont {Tang}}, \bibinfo
  {author} {\bibfnamefont {Kai}\ \bibnamefont {Chen}}, \bibinfo {author}
  {\bibfnamefont {Nobumichi}\ \bibnamefont {Tamura}}, \bibinfo {author}
  {\bibfnamefont {Phillip~M}\ \bibnamefont {Wu}}, \bibinfo {author}
  {\bibfnamefont {Jun}\ \bibnamefont {Fujioka}}, \bibinfo {author}
  {\bibfnamefont {Yoshinori}\ \bibnamefont {Tokura}}, \ and\ \bibinfo {author}
  {\bibfnamefont {Zhi-Xun}\ \bibnamefont {Shen}},\ }\bibfield  {title}
  {\enquote {\bibinfo {title} {{Mobile metallic domain walls in an
  all-in-all-out magnetic insulator}},}\ }\href
  {https://doi.org/10.1126/science.aac8289} {\bibfield  {journal} {\bibinfo
  {journal} {Science}\ }\textbf {\bibinfo {volume} {350}},\ \bibinfo {pages}
  {538--541} (\bibinfo {year} {2015})}\BibitemShut {NoStop}%
\bibitem [{\citenamefont {Tomiyasu}\ \emph {et~al.}(2012)\citenamefont
  {Tomiyasu}, \citenamefont {Matsuhira}, \citenamefont {Iwasa}, \citenamefont
  {Watahiki}, \citenamefont {Takagi}, \citenamefont {Wakeshima}, \citenamefont
  {Hinatsu}, \citenamefont {Yokoyama}, \citenamefont {Ohoyama},\ and\
  \citenamefont {Yamada}}]{tomiyasu2012neutrons}%
  \BibitemOpen
  \bibfield  {author} {\bibinfo {author} {\bibfnamefont {Keisuke}\ \bibnamefont
  {Tomiyasu}}, \bibinfo {author} {\bibfnamefont {Kazuyuki}\ \bibnamefont
  {Matsuhira}}, \bibinfo {author} {\bibfnamefont {Kazuaki}\ \bibnamefont
  {Iwasa}}, \bibinfo {author} {\bibfnamefont {Masanori}\ \bibnamefont
  {Watahiki}}, \bibinfo {author} {\bibfnamefont {Seishi}\ \bibnamefont
  {Takagi}}, \bibinfo {author} {\bibfnamefont {Makoto}\ \bibnamefont
  {Wakeshima}}, \bibinfo {author} {\bibfnamefont {Yukio}\ \bibnamefont
  {Hinatsu}}, \bibinfo {author} {\bibfnamefont {Makoto}\ \bibnamefont
  {Yokoyama}}, \bibinfo {author} {\bibfnamefont {Kenji}\ \bibnamefont
  {Ohoyama}}, \ and\ \bibinfo {author} {\bibfnamefont {Kazuyoshi}\ \bibnamefont
  {Yamada}},\ }\bibfield  {title} {\enquote {\bibinfo {title} {{Emergence of
  magnetic long-range order in frustrated pyrochlore Nd$_2$Ir$_2$O$_7$ with
  metal--insulator transition}},}\ }\href
  {https://doi.org/10.1143/JPSJ.81.034709} {\bibfield  {journal} {\bibinfo
  {journal} {Journal of the Physical Society of Japan}\ }\textbf {\bibinfo
  {volume} {81}},\ \bibinfo {pages} {034709} (\bibinfo {year}
  {2012})}\BibitemShut {NoStop}%
\bibitem [{\citenamefont {Ueda}\ \emph {et~al.}(2015)\citenamefont {Ueda},
  \citenamefont {Fujioka}, \citenamefont {Yang}, \citenamefont {Shiogai},
  \citenamefont {Tsukazaki}, \citenamefont {Nakamura}, \citenamefont {Awaji},
  \citenamefont {Nagaosa},\ and\ \citenamefont
  {Tokura}}]{ueda2015magnetoresistance}%
  \BibitemOpen
  \bibfield  {author} {\bibinfo {author} {\bibfnamefont {K.}~\bibnamefont
  {Ueda}}, \bibinfo {author} {\bibfnamefont {J.}~\bibnamefont {Fujioka}},
  \bibinfo {author} {\bibfnamefont {B.-J.}\ \bibnamefont {Yang}}, \bibinfo
  {author} {\bibfnamefont {J.}~\bibnamefont {Shiogai}}, \bibinfo {author}
  {\bibfnamefont {A.}~\bibnamefont {Tsukazaki}}, \bibinfo {author}
  {\bibfnamefont {S.}~\bibnamefont {Nakamura}}, \bibinfo {author}
  {\bibfnamefont {S.}~\bibnamefont {Awaji}}, \bibinfo {author} {\bibfnamefont
  {N.}~\bibnamefont {Nagaosa}}, \ and\ \bibinfo {author} {\bibfnamefont
  {Y.}~\bibnamefont {Tokura}},\ }\bibfield  {title} {\enquote {\bibinfo {title}
  {{Magnetic field-induced insulator-semimetal transition in a pyrochlore
  Nd$_2$Ir$_2$O$_7$}},}\ }\href
  {https://doi.org/10.1103/PhysRevLett.115.056402} {\bibfield  {journal}
  {\bibinfo  {journal} {Phys. Rev. Lett.}\ }\textbf {\bibinfo {volume} {115}},\
  \bibinfo {pages} {056402} (\bibinfo {year} {2015})}\BibitemShut {NoStop}%
\bibitem [{\citenamefont {Nakayama}\ \emph {et~al.}(2016)\citenamefont
  {Nakayama}, \citenamefont {Kondo}, \citenamefont {Tian}, \citenamefont
  {Ishikawa}, \citenamefont {Halim}, \citenamefont {Bareille}, \citenamefont
  {Malaeb}, \citenamefont {Kuroda}, \citenamefont {Tomita}, \citenamefont
  {Ideta}, \citenamefont {Tanaka}, \citenamefont {Matsunami}, \citenamefont
  {Kimura}, \citenamefont {Inami}, \citenamefont {Ono}, \citenamefont
  {Kumigashira}, \citenamefont {Balents}, \citenamefont {Nakatsuji},\ and\
  \citenamefont {Shin}}]{nakayama2016arpes}%
  \BibitemOpen
  \bibfield  {author} {\bibinfo {author} {\bibfnamefont {M.}~\bibnamefont
  {Nakayama}}, \bibinfo {author} {\bibfnamefont {Takeshi}\ \bibnamefont
  {Kondo}}, \bibinfo {author} {\bibfnamefont {Z.}~\bibnamefont {Tian}},
  \bibinfo {author} {\bibfnamefont {J.~J.}\ \bibnamefont {Ishikawa}}, \bibinfo
  {author} {\bibfnamefont {M.}~\bibnamefont {Halim}}, \bibinfo {author}
  {\bibfnamefont {C.}~\bibnamefont {Bareille}}, \bibinfo {author}
  {\bibfnamefont {W.}~\bibnamefont {Malaeb}}, \bibinfo {author} {\bibfnamefont
  {K.}~\bibnamefont {Kuroda}}, \bibinfo {author} {\bibfnamefont
  {T.}~\bibnamefont {Tomita}}, \bibinfo {author} {\bibfnamefont
  {S.}~\bibnamefont {Ideta}}, \bibinfo {author} {\bibfnamefont
  {K.}~\bibnamefont {Tanaka}}, \bibinfo {author} {\bibfnamefont
  {M.}~\bibnamefont {Matsunami}}, \bibinfo {author} {\bibfnamefont
  {S.}~\bibnamefont {Kimura}}, \bibinfo {author} {\bibfnamefont
  {N.}~\bibnamefont {Inami}}, \bibinfo {author} {\bibfnamefont
  {K.}~\bibnamefont {Ono}}, \bibinfo {author} {\bibfnamefont {H.}~\bibnamefont
  {Kumigashira}}, \bibinfo {author} {\bibfnamefont {L.}~\bibnamefont
  {Balents}}, \bibinfo {author} {\bibfnamefont {S.}~\bibnamefont {Nakatsuji}},
  \ and\ \bibinfo {author} {\bibfnamefont {S.}~\bibnamefont {Shin}},\
  }\bibfield  {title} {\enquote {\bibinfo {title} {{Slater to Mott crossover in
  the metal to insulator transition of Nd$_2$Ir$_2$O$_7$}},}\ }\href
  {https://doi.org/10.1103/PhysRevLett.117.056403} {\bibfield  {journal}
  {\bibinfo  {journal} {Phys. Rev. Lett.}\ }\textbf {\bibinfo {volume} {117}},\
  \bibinfo {pages} {056403} (\bibinfo {year} {2016})}\BibitemShut {NoStop}%
\bibitem [{\citenamefont {Tardif}\ \emph {et~al.}(2015)\citenamefont {Tardif},
  \citenamefont {Takeshita}, \citenamefont {Ohsumi}, \citenamefont {Yamaura},
  \citenamefont {Okuyama}, \citenamefont {Hiroi}, \citenamefont {Takata},\ and\
  \citenamefont {Arima}}]{tardif2015domains}%
  \BibitemOpen
  \bibfield  {author} {\bibinfo {author} {\bibfnamefont {Samuel}\ \bibnamefont
  {Tardif}}, \bibinfo {author} {\bibfnamefont {Soshi}\ \bibnamefont
  {Takeshita}}, \bibinfo {author} {\bibfnamefont {Hiroyuki}\ \bibnamefont
  {Ohsumi}}, \bibinfo {author} {\bibfnamefont {Jun-ichi}\ \bibnamefont
  {Yamaura}}, \bibinfo {author} {\bibfnamefont {Daisuke}\ \bibnamefont
  {Okuyama}}, \bibinfo {author} {\bibfnamefont {Zenji}\ \bibnamefont {Hiroi}},
  \bibinfo {author} {\bibfnamefont {Masaki}\ \bibnamefont {Takata}}, \ and\
  \bibinfo {author} {\bibfnamefont {Taka-hisa}\ \bibnamefont {Arima}},\
  }\bibfield  {title} {\enquote {\bibinfo {title} {{All-in--all-out magnetic
  domains: X-ray diffraction imaging and magnetic field control}},}\ }\href
  {https://doi.org/10.1103/PhysRevLett.114.147205} {\bibfield  {journal}
  {\bibinfo  {journal} {Phys. Rev. Lett.}\ }\textbf {\bibinfo {volume} {114}},\
  \bibinfo {pages} {147205} (\bibinfo {year} {2015})}\BibitemShut {NoStop}%
\bibitem [{\citenamefont {Witczak-Krempa}\ \emph {et~al.}(2013)\citenamefont
  {Witczak-Krempa}, \citenamefont {Go},\ and\ \citenamefont
  {Kim}}]{witczak2013effectiveHamiltonian}%
  \BibitemOpen
  \bibfield  {author} {\bibinfo {author} {\bibfnamefont {William}\ \bibnamefont
  {Witczak-Krempa}}, \bibinfo {author} {\bibfnamefont {Ara}\ \bibnamefont
  {Go}}, \ and\ \bibinfo {author} {\bibfnamefont {Yong~Baek}\ \bibnamefont
  {Kim}},\ }\bibfield  {title} {\enquote {\bibinfo {title} {{Pyrochlore
  electrons under pressure, heat, and field: Shedding light on the
  iridates}},}\ }\href {https://doi.org/10.1103/PhysRevB.87.155101} {\bibfield
  {journal} {\bibinfo  {journal} {Phys. Rev. B}\ }\textbf {\bibinfo {volume}
  {87}},\ \bibinfo {pages} {155101} (\bibinfo {year} {2013})}\BibitemShut
  {NoStop}%
\bibitem [{\citenamefont {Chen}\ and\ \citenamefont
  {Hermele}(2012)}]{chen2012fd}%
  \BibitemOpen
  \bibfield  {author} {\bibinfo {author} {\bibfnamefont {Gang}\ \bibnamefont
  {Chen}}\ and\ \bibinfo {author} {\bibfnamefont {Michael}\ \bibnamefont
  {Hermele}},\ }\bibfield  {title} {\enquote {\bibinfo {title} {{Magnetic
  orders and topological phases from f-d exchange in pyrochlore iridates}},}\
  }\href {https://doi.org/10.1103/PhysRevB.86.235129} {\bibfield  {journal}
  {\bibinfo  {journal} {Phys. Rev. B}\ }\textbf {\bibinfo {volume} {86}},\
  \bibinfo {pages} {235129} (\bibinfo {year} {2012})}\BibitemShut {NoStop}%
\bibitem [{\citenamefont {Pesin}\ and\ \citenamefont
  {Balents}(2010)}]{PesinBalents2010}%
  \BibitemOpen
  \bibfield  {author} {\bibinfo {author} {\bibfnamefont {Dmytro}\ \bibnamefont
  {Pesin}}\ and\ \bibinfo {author} {\bibfnamefont {Leon}\ \bibnamefont
  {Balents}},\ }\bibfield  {title} {\enquote {\bibinfo {title} {{Mott physics
  and band topology in materials with strong spin--orbit interaction}},}\
  }\href {https://doi.org/10.1038/nphys1606} {\bibfield  {journal} {\bibinfo
  {journal} {Nature Physics}\ }\textbf {\bibinfo {volume} {6}},\ \bibinfo
  {pages} {376--381} (\bibinfo {year} {2010})}\BibitemShut {NoStop}%
\bibitem [{\citenamefont {Zhu}\ \emph {et~al.}(2012)\citenamefont {Zhu},
  \citenamefont {Collaudin}, \citenamefont {Fauqu{\'e}}, \citenamefont {Kang},\
  and\ \citenamefont {Behnia}}]{zhu2012field}%
  \BibitemOpen
  \bibfield  {author} {\bibinfo {author} {\bibfnamefont {Zengwei}\ \bibnamefont
  {Zhu}}, \bibinfo {author} {\bibfnamefont {Aur{\'e}lie}\ \bibnamefont
  {Collaudin}}, \bibinfo {author} {\bibfnamefont {Beno{\^\i}t}\ \bibnamefont
  {Fauqu{\'e}}}, \bibinfo {author} {\bibfnamefont {Woun}\ \bibnamefont {Kang}},
  \ and\ \bibinfo {author} {\bibfnamefont {Kamran}\ \bibnamefont {Behnia}},\
  }\bibfield  {title} {\enquote {\bibinfo {title} {{Field-induced polarization
  of Dirac valleys in bismuth}},}\ }\href {https://doi.org/10.1038/nphys2111}
  {\bibfield  {journal} {\bibinfo  {journal} {Nature Physics}\ }\textbf
  {\bibinfo {volume} {8}},\ \bibinfo {pages} {89--94} (\bibinfo {year}
  {2012})}\BibitemShut {NoStop}%
\bibitem [{\citenamefont {Zhang}\ \emph {et~al.}(2017)\citenamefont {Zhang},
  \citenamefont {Haule},\ and\ \citenamefont {Vanderbilt}}]{Zhang2017LDA}%
  \BibitemOpen
  \bibfield  {author} {\bibinfo {author} {\bibfnamefont {Hongbin}\ \bibnamefont
  {Zhang}}, \bibinfo {author} {\bibfnamefont {Kristjan}\ \bibnamefont {Haule}},
  \ and\ \bibinfo {author} {\bibfnamefont {David}\ \bibnamefont {Vanderbilt}},\
  }\bibfield  {title} {\enquote {\bibinfo {title} {{Metal-Insulator Transition
  and Topological Properties of Pyrochlore Iridates}},}\ }\href
  {https://doi.org/10.1103/PhysRevLett.118.026404} {\bibfield  {journal}
  {\bibinfo  {journal} {Phys. Rev. Lett.}\ }\textbf {\bibinfo {volume} {118}},\
  \bibinfo {pages} {026404} (\bibinfo {year} {2017})}\BibitemShut {NoStop}%
\bibitem [{\citenamefont {Basov}\ \emph {et~al.}(2011)\citenamefont {Basov},
  \citenamefont {Averitt}, \citenamefont {van~der Marel}, \citenamefont
  {Dressel},\ and\ \citenamefont {Haule}}]{basov2011electrodynamics}%
  \BibitemOpen
  \bibfield  {author} {\bibinfo {author} {\bibfnamefont {Dimitri~N}\
  \bibnamefont {Basov}}, \bibinfo {author} {\bibfnamefont {Richard~D}\
  \bibnamefont {Averitt}}, \bibinfo {author} {\bibfnamefont {Dirk}\
  \bibnamefont {van~der Marel}}, \bibinfo {author} {\bibfnamefont {Martin}\
  \bibnamefont {Dressel}}, \ and\ \bibinfo {author} {\bibfnamefont {Kristjan}\
  \bibnamefont {Haule}},\ }\bibfield  {title} {\enquote {\bibinfo {title}
  {{Electrodynamics of correlated electron materials}},}\ }\href
  {https://doi.org/10.1103/RevModPhys.83.471} {\bibfield  {journal} {\bibinfo
  {journal} {Reviews of Modern Physics}\ }\textbf {\bibinfo {volume} {83}},\
  \bibinfo {pages} {471} (\bibinfo {year} {2011})}\BibitemShut {NoStop}%
\bibitem [{\citenamefont {Stricker}\ \emph {et~al.}(2014)\citenamefont
  {Stricker}, \citenamefont {Mravlje}, \citenamefont {Berthod}, \citenamefont
  {Fittipaldi}, \citenamefont {Vecchione}, \citenamefont {Georges},\ and\
  \citenamefont {van~der Marel}}]{stricker2014optical}%
  \BibitemOpen
  \bibfield  {author} {\bibinfo {author} {\bibfnamefont {Damien}\ \bibnamefont
  {Stricker}}, \bibinfo {author} {\bibfnamefont {Jernej}\ \bibnamefont
  {Mravlje}}, \bibinfo {author} {\bibfnamefont {Christophe}\ \bibnamefont
  {Berthod}}, \bibinfo {author} {\bibfnamefont {Rosalba}\ \bibnamefont
  {Fittipaldi}}, \bibinfo {author} {\bibfnamefont {Antonio}\ \bibnamefont
  {Vecchione}}, \bibinfo {author} {\bibfnamefont {Antoine}\ \bibnamefont
  {Georges}}, \ and\ \bibinfo {author} {\bibfnamefont {Dirk}\ \bibnamefont
  {van~der Marel}},\ }\bibfield  {title} {\enquote {\bibinfo {title} {{Optical
  response of Sr$_2$RuO$_4$ reveals universal fermi-liquid scaling and
  quasiparticles beyond Landau theory}},}\ }\href
  {https://doi.org/10.1103/PhysRevLett.113.087404} {\bibfield  {journal}
  {\bibinfo  {journal} {Phys. Rev. Lett.}\ }\textbf {\bibinfo {volume} {113}},\
  \bibinfo {pages} {087404} (\bibinfo {year} {2014})}\BibitemShut {NoStop}%
\bibitem [{\citenamefont {van Mechelen}\ \emph {et~al.}(2008)\citenamefont {van
  Mechelen}, \citenamefont {van~der Marel}, \citenamefont {Grimaldi},
  \citenamefont {Kuzmenko}, \citenamefont {Armitage}, \citenamefont {Reyren},
  \citenamefont {Hagemann},\ and\ \citenamefont {Mazin}}]{van2008electron}%
  \BibitemOpen
  \bibfield  {author} {\bibinfo {author} {\bibfnamefont {J.~L.~M.}\
  \bibnamefont {van Mechelen}}, \bibinfo {author} {\bibfnamefont
  {D.}~\bibnamefont {van~der Marel}}, \bibinfo {author} {\bibfnamefont
  {C.}~\bibnamefont {Grimaldi}}, \bibinfo {author} {\bibfnamefont {A.~B.}\
  \bibnamefont {Kuzmenko}}, \bibinfo {author} {\bibfnamefont {N.~P.}\
  \bibnamefont {Armitage}}, \bibinfo {author} {\bibfnamefont {N.}~\bibnamefont
  {Reyren}}, \bibinfo {author} {\bibfnamefont {H.}~\bibnamefont {Hagemann}}, \
  and\ \bibinfo {author} {\bibfnamefont {I.~I.}\ \bibnamefont {Mazin}},\
  }\bibfield  {title} {\enquote {\bibinfo {title} {{Electron-phonon interaction
  and charge carrier mass enhancement in SrTiO$_3$}},}\ }\href
  {https://doi.org/10.1103/PhysRevLett.100.226403} {\bibfield  {journal}
  {\bibinfo  {journal} {Phys. Rev. Lett.}\ }\textbf {\bibinfo {volume} {100}},\
  \bibinfo {pages} {226403} (\bibinfo {year} {2008})}\BibitemShut {NoStop}%
\bibitem [{\citenamefont {Moon}\ \emph {et~al.}(2008)\citenamefont {Moon},
  \citenamefont {Jin}, \citenamefont {Kim}, \citenamefont {Choi}, \citenamefont
  {Lee}, \citenamefont {Yu}, \citenamefont {Cao}, \citenamefont {Sumi},
  \citenamefont {Funakubo}, \citenamefont {Bernhard},\ and\ \citenamefont
  {Noh}}]{moon2008SrIrO}%
  \BibitemOpen
  \bibfield  {author} {\bibinfo {author} {\bibfnamefont {S.~J.}\ \bibnamefont
  {Moon}}, \bibinfo {author} {\bibfnamefont {H.}~\bibnamefont {Jin}}, \bibinfo
  {author} {\bibfnamefont {K.~W.}\ \bibnamefont {Kim}}, \bibinfo {author}
  {\bibfnamefont {W.~S.}\ \bibnamefont {Choi}}, \bibinfo {author}
  {\bibfnamefont {Y.~S.}\ \bibnamefont {Lee}}, \bibinfo {author} {\bibfnamefont
  {J.}~\bibnamefont {Yu}}, \bibinfo {author} {\bibfnamefont {G.}~\bibnamefont
  {Cao}}, \bibinfo {author} {\bibfnamefont {A.}~\bibnamefont {Sumi}}, \bibinfo
  {author} {\bibfnamefont {H.}~\bibnamefont {Funakubo}}, \bibinfo {author}
  {\bibfnamefont {C.}~\bibnamefont {Bernhard}}, \ and\ \bibinfo {author}
  {\bibfnamefont {T.~W.}\ \bibnamefont {Noh}},\ }\bibfield  {title} {\enquote
  {\bibinfo {title} {{Dimensionality-Controlled Insulator-Metal Transition and
  Correlated Metallic State in 5 d Transition Metal Oxides
  Sr$_{n+1}$Ir$_n$O$_{3n+1}$ (n= 1, 2, and $\infty$)}},}\ }\href
  {https://doi.org/10.1103/PhysRevLett.101.226402} {\bibfield  {journal}
  {\bibinfo  {journal} {Phys. Rev. Lett.}\ }\textbf {\bibinfo {volume} {101}},\
  \bibinfo {pages} {226402} (\bibinfo {year} {2008})}\BibitemShut {NoStop}%
\bibitem [{\citenamefont {Wang}\ \emph {et~al.}(2018)\citenamefont {Wang},
  \citenamefont {Bachar}, \citenamefont {Teyssier}, \citenamefont {Luo},
  \citenamefont {Rischau}, \citenamefont {Scheerer}, \citenamefont {de~la
  Torre}, \citenamefont {Perry}, \citenamefont {Baumberger},\ and\
  \citenamefont {van~der Marel}}]{wang2018mott}%
  \BibitemOpen
  \bibfield  {author} {\bibinfo {author} {\bibfnamefont {K.}~\bibnamefont
  {Wang}}, \bibinfo {author} {\bibfnamefont {N.}~\bibnamefont {Bachar}},
  \bibinfo {author} {\bibfnamefont {J.}~\bibnamefont {Teyssier}}, \bibinfo
  {author} {\bibfnamefont {W.}~\bibnamefont {Luo}}, \bibinfo {author}
  {\bibfnamefont {C.~W.}\ \bibnamefont {Rischau}}, \bibinfo {author}
  {\bibfnamefont {G.}~\bibnamefont {Scheerer}}, \bibinfo {author}
  {\bibfnamefont {A.}~\bibnamefont {de~la Torre}}, \bibinfo {author}
  {\bibfnamefont {R.~S.}\ \bibnamefont {Perry}}, \bibinfo {author}
  {\bibfnamefont {F.}~\bibnamefont {Baumberger}}, \ and\ \bibinfo {author}
  {\bibfnamefont {D.}~\bibnamefont {van~der Marel}},\ }\bibfield  {title}
  {\enquote {\bibinfo {title} {{Mott transition and collective charge pinning
  in electron doped Sr$_2$IrO$_4$}},}\ }\href
  {https://doi.org/10.1103/PhysRevB.98.045107} {\bibfield  {journal} {\bibinfo
  {journal} {Phys. Rev. B}\ }\textbf {\bibinfo {volume} {98}},\ \bibinfo
  {pages} {045107} (\bibinfo {year} {2018})}\BibitemShut {NoStop}%
\bibitem [{\citenamefont {Shinaoka}\ \emph {et~al.}(2019)\citenamefont
  {Shinaoka}, \citenamefont {Motome}, \citenamefont {Miyake}, \citenamefont
  {Ishibashi},\ and\ \citenamefont {Werner}}]{shinaoka2019DFT}%
  \BibitemOpen
  \bibfield  {author} {\bibinfo {author} {\bibfnamefont {Hiroshi}\ \bibnamefont
  {Shinaoka}}, \bibinfo {author} {\bibfnamefont {Yukitoshi}\ \bibnamefont
  {Motome}}, \bibinfo {author} {\bibfnamefont {Takashi}\ \bibnamefont
  {Miyake}}, \bibinfo {author} {\bibfnamefont {Shoji}\ \bibnamefont
  {Ishibashi}}, \ and\ \bibinfo {author} {\bibfnamefont {Philipp}\ \bibnamefont
  {Werner}},\ }\bibfield  {title} {\enquote {\bibinfo {title}
  {{First-principles studies of spin-orbital physics in pyrochlore oxides}},}\
  }\href {https://doi.org/10.1088/1361-648x/ab162f} {\bibfield  {journal}
  {\bibinfo  {journal} {Journal of Physics: Condensed Matter}\ }\textbf
  {\bibinfo {volume} {31}},\ \bibinfo {pages} {323001} (\bibinfo {year}
  {2019})}\BibitemShut {NoStop}%
\bibitem [{\citenamefont {Kondo}\ \emph {et~al.}(2015)\citenamefont {Kondo},
  \citenamefont {Nakayama}, \citenamefont {Chen}, \citenamefont {Ishikawa},
  \citenamefont {Moon}, \citenamefont {Yamamoto}, \citenamefont {Ota},
  \citenamefont {Malaeb}, \citenamefont {Kanai}, \citenamefont {Nakashima},
  \citenamefont {Ishida}, \citenamefont {Yoshida}, \citenamefont {Yamamoto},
  \citenamefont {Matsunami}, \citenamefont {Inami}, \citenamefont {Ono},
  \citenamefont {Kumigashira}, \citenamefont {Nakatsuji}, \citenamefont
  {Balents},\ and\ \citenamefont {Shin}}]{kondo2015arpes}%
  \BibitemOpen
  \bibfield  {author} {\bibinfo {author} {\bibfnamefont {Takeshi}\ \bibnamefont
  {Kondo}}, \bibinfo {author} {\bibfnamefont {M.}~\bibnamefont {Nakayama}},
  \bibinfo {author} {\bibfnamefont {R.}~\bibnamefont {Chen}}, \bibinfo {author}
  {\bibfnamefont {J.~J.}\ \bibnamefont {Ishikawa}}, \bibinfo {author}
  {\bibfnamefont {E.-G.}\ \bibnamefont {Moon}}, \bibinfo {author}
  {\bibfnamefont {T.}~\bibnamefont {Yamamoto}}, \bibinfo {author}
  {\bibfnamefont {Y.}~\bibnamefont {Ota}}, \bibinfo {author} {\bibfnamefont
  {W.}~\bibnamefont {Malaeb}}, \bibinfo {author} {\bibfnamefont
  {H.}~\bibnamefont {Kanai}}, \bibinfo {author} {\bibfnamefont
  {Y.}~\bibnamefont {Nakashima}}, \bibinfo {author} {\bibfnamefont
  {Y.}~\bibnamefont {Ishida}}, \bibinfo {author} {\bibfnamefont
  {R.}~\bibnamefont {Yoshida}}, \bibinfo {author} {\bibfnamefont
  {H.}~\bibnamefont {Yamamoto}}, \bibinfo {author} {\bibfnamefont
  {S.}~\bibnamefont {Matsunami}, \bibfnamefont {M. and~Kimura}}, \bibinfo
  {author} {\bibfnamefont {N.}~\bibnamefont {Inami}}, \bibinfo {author}
  {\bibfnamefont {K.}~\bibnamefont {Ono}}, \bibinfo {author} {\bibfnamefont
  {H.}~\bibnamefont {Kumigashira}}, \bibinfo {author} {\bibfnamefont
  {S.}~\bibnamefont {Nakatsuji}}, \bibinfo {author} {\bibfnamefont { ~L.}\
  \bibnamefont {Balents}}, \ and\ \bibinfo {author} {\bibfnamefont
  {S.}~\bibnamefont {Shin}},\ }\bibfield  {title} {\enquote {\bibinfo {title}
  {{Quadratic Fermi node in a 3D strongly correlated semimetal}},}\ }\href
  {https://doi.org/10.1038/ncomms10042} {\bibfield  {journal} {\bibinfo
  {journal} {Nature Communications}\ }\textbf {\bibinfo {volume} {6}},\
  \bibinfo {pages} {1--8} (\bibinfo {year} {2015})}\BibitemShut {NoStop}%
\bibitem [{\citenamefont {Wang}\ \emph {et~al.}(2017)\citenamefont {Wang},
  \citenamefont {Go},\ and\ \citenamefont {Millis}}]{wang2017weyl}%
  \BibitemOpen
  \bibfield  {author} {\bibinfo {author} {\bibfnamefont {Runzhi}\ \bibnamefont
  {Wang}}, \bibinfo {author} {\bibfnamefont {Ara}\ \bibnamefont {Go}}, \ and\
  \bibinfo {author} {\bibfnamefont {Andrew}\ \bibnamefont {Millis}},\
  }\bibfield  {title} {\enquote {\bibinfo {title} {{Weyl rings and enhanced
  susceptibilities in pyrochlore iridates: k{\textperiodcentered} p analysis of
  cluster dynamical mean-field theory results}},}\ }\href
  {https://doi.org/10.1103/PhysRevB.96.195158} {\bibfield  {journal} {\bibinfo
  {journal} {Phys. Rev. B}\ }\textbf {\bibinfo {volume} {96}},\ \bibinfo
  {pages} {195158} (\bibinfo {year} {2017})}\BibitemShut {NoStop}%
\bibitem [{\citenamefont {Ashby}\ and\ \citenamefont
  {Carbotte}(2014)}]{ashby2014chiral}%
  \BibitemOpen
  \bibfield  {author} {\bibinfo {author} {\bibfnamefont {Phillip E.~C.}\
  \bibnamefont {Ashby}}\ and\ \bibinfo {author} {\bibfnamefont {J.~P.}\
  \bibnamefont {Carbotte}},\ }\bibfield  {title} {\enquote {\bibinfo {title}
  {{Chiral anomaly and optical absorption in Weyl semimetals}},}\ }\href
  {https://doi.org/10.1103/PhysRevB.89.245121} {\bibfield  {journal} {\bibinfo
  {journal} {Phys. Rev. B}\ }\textbf {\bibinfo {volume} {89}},\ \bibinfo
  {pages} {245121} (\bibinfo {year} {2014})}\BibitemShut {NoStop}%
\end{thebibliography}
%
\end{document}